\documentclass[usegraphicx,usenatbib]{mn2e}
\usepackage{color,times,float}

\newcommand{\Msol}{\ensuremath{M_\odot}}
\newcommand{\code}[1]{\textsc{#1}}

\newcommand*{\rom}[1]{\romannumeral #1}
\newcommand{\ion}[2]{\textrm{#1}~\textsc{\rom{#2}}}

\newcommand{\nickel}{{\ensuremath{^{56}\mathrm{Ni}}}}
\newcommand{\cobalt}{{\ensuremath{^{56}\mathrm{Co}}}}
\newcommand{\iron}{{\ensuremath{^{56}\mathrm{Fe}}}}

\newcommand{\vSi}{\ensuremath{v_{\mathrm{Si}}}}

\newcommand{\kms}{\ensuremath{\mathrm{km~s}^{-1}}}

\newcommand{\MNi}{\ensuremath{M_{\mathrm{Ni}}}}
\newcommand{\Mej}{\ensuremath{M_\mathrm{ej}}}
\newcommand{\Mch}{\ensuremath{M_\mathrm{Ch}}}
\newcommand{\vKE}{\ensuremath{v_\mathrm{KE}}}
\newcommand{\aNi}{\ensuremath{a_\mathrm{Ni}}}
\newcommand{\kgamma}{\ensuremath{\kappa_\gamma}}

\newcommand{\Bmax}{{\ensuremath{B_\mathrm{max}}}}
\newcommand{\sBV}{{\ensuremath{s_{BV}}}}
\newcommand{\ebmvhost}{\mbox{\ensuremath{E(B-V)_\mathrm{host}}}}
\newcommand{\rvhost}{\mbox{\ensuremath{R_{V,\mathrm{host}}}}}
\newcommand{\snoopy}{\code{SNooPy}}
\newcommand{\bayessn}{\code{BayeSN}}
\newcommand{\salt}{\code{SALT2}}
\newcommand{\snid}{\code{SNID}}
\newcommand{\bolomass}{\code{bolomass}}
\newcommand{\emcee}{\code{emcee}}
\newcommand{\Lbol}[1]{\ensuremath{L_\mathrm{{#1},bol}}}
\newcommand{\dmbol}[1]{\ensuremath{\Delta m_\mathrm{{#1},bol}}}
\renewcommand{\vec}[1]{\ensuremath{\bf {#1}}}
\newcommand{\csp}{CSP-I}

\definecolor{kellygreen}{rgb}{0.3, 0.73, 0.09}
\newcommand{\revIIb}[1]{{{#1}}}
\newcommand{\revIIg}[1]{{{#1}}}
\newcommand{\nb}[1]{\ensuremath{^{#1}}}

\begin{document}


\title[CSP-I + CfA bolometric light curves]
      {Probing type Ia supernova properties using bolometric light curves
       from the Carnegie Supernova Project and the CfA Supernova Group}

\author[Scalzo et al.]
{
    R.~A.~Scalzo\nb{1,2,3}\thanks{Email: rscalzo@mso.anu.edu.au},
    E.~Parent\nb{4,5},
    C. Burns\nb{4},
    M.~Childress\nb{1,3},
    B.~E.~Tucker\nb{1,3,6}, \newauthor
    P.~J.~Brown\nb{7},
    C.~Contreras\nb{8},
    E.~Hsiao\nb{8},
    K.~Krisciunas\nb{7},
    N.~Morrell\nb{8},
    M.~M.~Phillips\nb{8},\newauthor
    A.~L.~Piro\nb{8},
    M.~Stritzinger\nb{9},
    and N.~Suntzeff\nb{7} \newauthor
    \\
    \nb{1} Research School of Astronomy and Astrophysics,
           Australian National University,
           Canberra, ACT 2611, Australia \\
    \nb{2} Centre for Translational Data Science, University of Sydney,
           Darlington, NSW 2008, Australia \\
    \nb{3} ARC Centre of Excellence for All-Sky Astrophysics (CAASTRO) \\
    \nb{4} Observatories of the Carnegie Institution for Science,
           Pasadena, CA 91101, USA \\
    \nb{5} Department of Physics and McGill Space Institute,
           McGill University, Montr\'eal, QC Canada H3A 2T8, Canada \\
    \nb{6} Department of Astronomy, University of California, Berkeley,
           B-20 Hearst Field Annex \#3411, Berkeley, CA 94720-3411, USA \\
    \nb{7} George P. and Cynthia Woods Mitchell Institute
           for Fundamental Physics and Astronomy,
           Department of Physics and Astronomy, Texas A\&M University, \\
           4242 TAMU, College Station, TX 77843, USA \\
    \nb{8} Carnegie Observatories, Las Campanas Observatory,
            La Serena, Chile \\
    \nb{9} Department of Physics and Astronomy, Aarhus University,
           Ny Munkegade 120, DK-8000 Aarhus C, Denmark \\
}

\maketitle

\vspace{-1in}

\begin{abstract}

We present bolometric light curves constructed from multi-wavelength
photometry of Type Ia supernovae (SNe~Ia) from the Carnegie Supernova Project
and the CfA Supernova Group, using near-infrared observations to provide
robust constraints on host galaxy dust extinction.  This set of light curves
form a well-measured reference set for comparison with theoretical models.
Ejected mass and synthesized \nickel\ mass are inferred for each SN~Ia from
its bolometric light curve using a semi-analytic Bayesian light curve model,
and fitting formulae provided in terms of light curve width parameters
from the \code{salt2} and \snoopy\ light curve fitters.
A weak bolometric width-luminosity relation is confirmed,
along with a correlation between ejected mass and the bolometric
light curve width.
{SNe~Ia likely to have sub-Chandrasekhar ejected masses}
belong preferentially to
the broad-line and cool-photosphere spectroscopic subtypes, and have higher
photospheric velocities and populate older, higher-mass host galaxies than
SNe~Ia consistent with Chandrasekhar-mass explosions.
Two peculiar events, SN~2006bt and SN~2006ot, have normal
peak luminosities but {appear to have} super-Chandrasekhar ejected
masses.
\end{abstract}



\vspace{0.1in}

\section{Introduction}

Type Ia supernovae (SNe~Ia), the thermonuclear explosions of white dwarfs,
were used as extragalactic distance indicators in the discovery of the
Universe's accelerating expansion \citep{riess98,perlmutter99}.
{They play a leading role in ongoing studies aimed at measuring
the Hubble constant and the nature of dark energy, and are also a critical
component to} the chemical enrichment of galaxies over cosmic time
\citep{kn09}.  Despite their central role in astrophysics and cosmology,
the evolutionary channels leading to the explosion and the final
explosion trigger for SNe~Ia have not yet been unambiguously identified;
this represents a challenging, long-standing unsolved problem in the field
\citep[for in-depth reviews, see:][]{wh12,hillebrandt13,pilar14}.

\revIIg{Accurate distance measurements} to SNe~Ia depend on empirical
relations between SN~Ia peak luminosity, light curve width, and color
\citep{phillips93,phillips99,riess96,tripp98,goldhaber01,guy07}
and, more recently, {a correction based on the} host galaxy mass
\citep{kelly10,sullivan11,childress13b}.  There are also established
relations between SN~Ia luminosity and temperature-dependent ratios
of spectral features \citep{nugent95,bongard06,silverman12b}, themselves
correlated with the light curve width.  These relations are
well-established observationally, and represent strong constraints on SN~Ia
explosion physics which still remain to be fully explained theoretically.
Identification of the SN~Ia progenitors could drive theoretical searches
for \revIIg{new}, independent luminosity correlates, decreasing statistical
and systematic uncertainties in measurements of the cosmic distance scale
and expansion history.

In most scenarios, the explosion is triggered by
interaction of the white dwarf with a binary companion, either a
non-degenerate star \citep[``single-degenerate'';][]{wi73} or another white
dwarf \citep[``double-degenerate'';][]{it84}.  \revIIg{In the traditional
single-degenerate scenario, a carbon-oxygen white dwarf accretes hydrogen
from its companion until igniting spontaneously near the Chandrasekhar
limiting mass $\Mch = 1.4~\Msol$; tests of SN~Ia progenitor scenarios thus
often focus on evidence for circumstellar hydrogen or for accretion processes.
Direct searches for surviving companions in SN~Ia remnants \citep{sp12}
and in pre-explosion imaging \citep{li11} yielded null results.
Upper limits on ionizing radiation from nuclear burning of accreted material
on the white dwarf's surface \citep{gb10,wg13,wg14} constrain accretion rates.
Upper limits on circumstellar hydrogen in most SN~Ia systems
come from non-detections of interaction flux in early light curves}
\citep{hayden10b,nugent11,bloom12,olling15,shappee15};
H$\alpha$ in late-time spectra \citep{mattila05,leonard07,shappee13};
radio emission \citep{panagia06,chomiuk12,chomiuk16};
and X-ray emission \citep{margutti14}.  \revIIg{However,}
these limits strictly rule out only symbiotic nova systems with red
giant companion stars.  \revIIg{The peculiar ``Ia-CSM''
\citep{silverman13} subclass shows significant circumstellar interaction
luminosity and narrow H$\alpha$ emission
\citep{hamuy03,mwv04,aldering06,prieto07,dilday12,taddia12},
but this subclass comprises at most a few percent of all SNe~Ia.}
\citet{cao15}, \citet{marion16},
{and \citet{hoss17}} present evidence for signatures of
single-degenerate companions in the near-ultraviolet and optical light curves
of otherwise normal SNe~Ia, starting within the first day after explosion
{but see \citep[but see][for conflicting evidence in one case]
{shappee18}}.
\citet{jiang17} report photometric and spectroscopic signatures
of \revIIg{\emph{helium} accretion} onto the white dwarf that produced the
well-observed SN~Ia MUSSES1604D.
{Nucleosynthetic constraints, sensitive to the central density of the
exploding white dwarf, may also present evidence for or against particular
progenitor scenarios \citep[e.g.][]{seitenzahl13b,mcwilliam17,shappee17}}.

Another approach is to compare observations to detailed
computational models of SN~Ia explosions
{
\citep{fritz12,blondin12,blondin13,diemer13,blondin17,hoeflich17,goldstein18}}.
These comparisons typically focus on tell-tale spectroscopic features or
light curves in specific passbands,
{and/or the distribution of \nickel\ in the ejecta.}
However, the radiation transfer problem for SN~Ia atmospheres
remains extremely challenging, and any practical solution will approximate
some aspects of the physics.  Full reproduction of the spectrum, including
velocities, strengths, and detailed shapes of atomic line features,
is the most stringent possible end-to-end test of an explosion model
\citep{dessart14a,dessart14b}.  It can be difficult to determine whether
discrepancies with observations represent failures of the underlying scenario
or merely some aspect of the calculation.  Single strong spectroscopic
features with uncertain behavior, such as the \ion{Ca}{2} infrared triplet
\citep{kasen06}, may also have dramatic influence on single-band light curves.

In contrast, the \emph{bolometric} light curve
--- the total radiant energy output from the SN~Ia as a function of time
--- is easier to simulate \revIIg{numerically
{\citep[e.g.][]{wygoda17,sukhbold18}}}
and can even be predicted semi-analytically \citep{arnett82,pe00a,pe00b},
but requires high-quality data with
broad wavelength and {high-cadence} temporal coverage to measure
observationally.  The bolometric light curve is sensitive to fundamental
physical parameters of the explosion, including the mass \MNi\ of radioactive
\nickel\ synthesized (which powers the light curve via the decay chain
\mbox{$\nickel \rightarrow \cobalt \rightarrow \iron$}) and the total ejected
mass \Mej.  These global parameters provide a complementary probe of the
different explosion mechanisms currently being tested.
\revIIg{The traditional single-degenerate scenario implies $\Mej = \Mch$, but}
other mechanisms with different progenitor masses could produce events
resembling SNe~Ia:  explosions of rapidly rotating,
super-Chandrasekhar-mass white dwarfs partially supported by accreted angular
momentum \citep{justham11,rds12}; super-Chandrasekhar-mass mergers of two
white dwarfs \citep[e.g.][]{pakmor11,pakmor12}; ``tamped detonations''
resulting from relaxed white dwarf merger products surrounded by a thick
carbon-oxygen envelope \citep{kmh93,hk96}; \emph{helium} detonations on
a sub-Chandrasekhar-mass white dwarf's surface \citep{ww94,sim10,fink10};
and collisions of two white dwarfs
\citep{rosswog09,raskin10,thompson11,kushnir13}.
Some authors have even looked into mechanisms enabling the spontaneous
explosion of isolated white dwarfs \citep{chiosi15,bramante15}.

Full bolometric light curves have been built for a relatively small sample
of normal SNe~Ia \citep{suntzeff96,vl96,contardo00,stritz06,scalzo14a}.
{However, the available light curves have had a significant impact
on development of SN~Ia theory.  \citet{stritz06} used semi-analytic
modeling of 16 SN~Ia bolometric light curves to infer a range of ejected
masses from 0.5--1.4~\Msol, which spurred important advances into
sub-Chandrasekhar-mass explosion models \citep{fink10,kromer10,sim10}.
\citet{scalzo14a} used an improved version of the technique on 19 additional
SNe~Ia in the nearby Hubble flow, finding} evidence that most normal SNe~Ia
\revIIg{have $\Mej = $1.0--1.4~\Msol, and that \Mej}\ correlates strongly
with light curve width parameters used in cosmology \citep{scalzo14a}.
\citet{scalzo14c} used this correlation as a starting point to reconstruct the
intrinsic distribution of \revIIg{\Mej}\ from a much larger sample of SNe~Ia;
they found that a significant fraction (at least 25\%) of all normal SNe~Ia
must explode beneath the Chandrasekhar limiting mass for white dwarfs,
and that the joint \Mej-\MNi\ distribution
could not be explained by any single contemporary explosion scenario.
{Subsequent theoretical work supports connections between the mass
distribution of SNe~Ia and the width-luminosity relation
\citep{wygoda17,blondin17,blondin18,goldstein18}}.

This work presents a set of high-quality bolometric light curves
constructed with public data from the \emph{Carnegie Supernova Project}
\citep[\csp;][]{hamuy06}
and the Harvard-Smithsonian Center for Astrophysics Supernova Group (CfA).
Sample selection is described in \S\ref{sec:obs}, host galaxy reddening
in \S\ref{sec:extinction}, and the procedure
for constructing bolometric light curves from multi-band data in
\S\ref{sec:bolo}.  A semianalytic modeling suite
(\S\ref{sec:bolomass}) developed in previous papers
\citep{scalzo10,scalzo12,scalzo14a,scalzo14b}
is used to infer \nickel\ masses and ejected masses from the bolometric
light curves.  Correlations between these global explosion parameters and
other observables such as spectroscopic subtype or host galaxy mass
are presented in \S\ref{sec:results}.
Implications for the width-luminosity relation, the physics
of peculiar SNe~Ia, and related questions are examined in
\S\ref{sec:discussion}, and conclusions and prospects for future work set out
in \S\ref{sec:conclusions}.


\section{Observations and Sample Selection}
\label{sec:obs}

The SNe~Ia we use for our investigation are drawn from the multi-year \csp\
and CfA data sets.  \revIIg{These surveys followed targets from searches that
target known nearby galaxies, unlike the untargeted search and follow-up
program by the Nearby Supernova Factory that
discovered most of the SNe~Ia analyzed in \citet{scalzo14a}.  Peculiar events
are over-represented since they are strongly selected for follow-up
observations;
they will be brighter and easier to observe if they are overluminous,
and will be observed more aggressively than normal SNe~Ia if they are
subluminous.  The current sample is thus useful for exploring the diversity
of SN~Ia bolometric light curve behavior.}

We use \csp\ $uBVgri$ photometry from \citet{cspdr2},
and CfA $UBVRIr'i'$ photometry from \citet{jha07}, \citet{cfa3},
and \citet{cfa4}.  Near-infrared (NIR) photometry is also available from
\citet{cspdr2} for \csp\ targets, and from \citet{cfair2} for CfA targets.
We use all photometry in {each group's} natural system, based on
the transmission functions published in the source papers.

We use derived spectroscopic quantities published in \citet{blondin12},
\citet{silverman12b}, and \citet{folatelli13}, including:
\begin{enumerate}
\item the heliocentric and CMB-frame redshifts of the host galaxy;
\item the blueshift velocity \vSi\ of the \ion{Si}{2}~$\lambda$6355 feature
      in spectra near maximum light;
\item the spectroscopic subtype identified by \snid\ \citep{snid}:
\item the \citet{wang09} subtype --- ``normal'' (N) or ``high-velocity'' (HV)
      --- determined by whether \vSi\ is greater or less than 11,800~\kms;
\item the \citet{branch06} subtype, based on measurements of
      the equivalent widths and line profile shapes of the
      \ion{Si}{2}~$\lambda$5972 and \ion{Si}{2}~$\lambda$6355 features
      in spectra taken near maximum light.
\end{enumerate}

Some SNe~Ia in our sample have host galaxy stellar masses available from
previous literature analyses.  Twenty-two targets have host masses from
\citet{neill09}, who employ common-aperture photometry on multi-wavelength
data for the host galaxies of nearby targets.  Similarly, \citet{childress13a}
derive host masses with common aperture photometry for the sample of SNe~Ia
observed by the Nearby Supernova Factory; 4 SNe~Ia from our sample have
published host galaxy masses from this work.  Finally, 2 SNe~Ia from our
sample have host masses from \citet{kelly10}.  All of these analyses have
comparable mass values (i.e. compatible initial mass functions).  We adopt
errors on host mass values as the quadrature sum of the published mass errors
(from measurement errors) and a 0.15~dex systematic error term which
\citet{childress13a} found to be an appropriate assessment of the systematic
uncertainty on galaxy mass-to-light ratios arising from variations in
star-formation histories.

The degree of temporal and wavelength completeness required for building
broad-band bolometric light curves means that even some of the best-observed
objects may be missing data in observationally demanding bandpasses
such as NIR.  To minimize the impact of corrections for missing flux,
we apply strict selection criteria described below, starting from
a total of 358 targets (85 \revIIg{from} \csp\ and 324 \revIIg{from} CfA,
with 34 observed by both programs).

An accurate measurement of a SN~Ia's luminosity requires an accurate distance
measurement, which can be ensured by restricting the sample to
the smooth Hubble flow, since direct distance measurements
are in general not available for very nearby targets.  However, most of the
\csp\ and CfA SNe~Ia were discovered by searches targeting specific nearby
galaxies, and closer SNe~Ia will in general have higher-quality,
more complete data.  To avoid too strict a selection, we choose targets
with $z > 0.013$ (4000~\kms).  Assuming a random peculiar velocity
of 300~\kms\ for each SN~Ia host galaxy {\citep{davis11}},
the induced systematic error on the
peak luminosity of each SN~Ia, and hence the \nickel\ mass derived from the
light curve, will be less than 15\% --- about the limit of accuracy that can
be achieved with the inference methods of \citet{scalzo14a} with the best
available data.  This cut removes 16 \csp\ targets and 62 CfA targets.

To capture as much of the SN~Ia radiation as possible at each observation
epoch, and to adequately sample the shape of the bolometric light curve,
we require each target to have the equivalent of full wavelength coverage
from 4000--9000~\AA\ ($BVRI$ equivalent) for at least one time point from
each of a set of key light curve phases, defined with respect to the date
of $B$-band maximum light as defined by the ``color model'' of the \snoopy\
light curve fitter \citep{burns11,burns14}:
\begin{enumerate}
\item between days $-8$ and $+0$ ($BV$ bands only), to ensure a robust
      constraint on the light curve width and host galaxy reddening;
\item within 3 days of day~$-1$ \citep[bolometric maximum;][]{scalzo14a},
      to ensure a robust constraint on the \nickel\ mass;
\item within 3 days of day~$+14$, to ensure that the decline of
      the bolometric light curve is well-constrained;
\item between day~$+21$ and day~$+35$, to ensure a robust constraint on
      the evolution of the spectral energy distribution (SED)
      between photospheric and early nebular phase; and
\item between day~$+40$ and day~$+80$, to constrain the late-time light
      curve and the ejected mass.
\end{enumerate}
Our best targets will also have 3300--4000~\AA\ ($U$ or $u$ equivalent)
at all of these epochs, enabling measurement of the full $UBVRI$ flux.
Other targets have good $U$/$u$ coverage near maximum light,
but with deteriorating signal-to-noise post-maximum, as line
blanketing from developing \ion{Fe}{2} features redistributes flux from blue
wavelengths to the NIR.  The requirement of $U$-band or $u$-band coverage
at more than three weeks past maximum light is thus restrictive,
in tension with a required minimum redshift for our targets.
The $U$/$u$ light curves vary significantly between individual SNe~Ia,
but the \csp\ $u-g$ and CfA $U-B$ colors evolve slowly after day~$+20$,
and the contribution to the bolometric flux could be adequately modeled
by a template at these late phases.  We therefore require $U/u$ data only
out to day~$+20$.

The phase coverage requirements eliminate 40 \csp\ targets and 249 CfA targets.
We are left with 39 unique SNe~Ia:  29 with \csp\ data, 13 with CfA data, and 3
(SN~2005eq, SN~2005hc, and SN~2006ax) with light curves from both programs
that pass all of our selection criteria.
Of these, 27 targets have $U/u$ coverage out past day~$+40$.

Our sample includes the spectroscopic subtype exemplar SN~1999aa,
the slow-declining SN~2004gu, and the 1991bg-like SN~2006gt and SN~2007ba.
It also includes the CfA light curve of the peculiar SN~Ia~2006bt
\citep{foley10}.  To examine this interesting target in more detail, we also
include the \csp\ light curves of SN~2006bt and the similar event
SN~2006ot \citep[from][]{cspdr2}, which have excellent temporal and wavelength
coverage across the region critical for our analysis but do not pass all
of our formal selection criteria.  We consider in detail the impact of missing
data and photometric peculiarity in our analysis of these events.
Finally, we include the CfA light curve of the
overluminous ``super-Chandra'' SN~2006gz \citep{hicken07}, which has never
undergone this type of detailed bolometric light curve analysis
despite its importance in {characterizing the observational
properties of this subclass.}  SN~2006gz lacks full
wavelength coverage around day~$+14$, but this will not affect our inference
of explosion properties.

Whenever possible, we have used NIR photometry to improve
constraints on the reddening \ebmvhost\ and the extinction law slope \rvhost\
due to dust in the host galaxy.  The NIR behavior is quite regular, with its
contribution to the luminosity ranging from $6\%$ near $B$-band maximum light
to nearly $30\%$ a few weeks later, and can be well-modeled by a template
\citep{scalzo14a}.
Most of our targets have at least some NIR data near maximum light.
Nine targets also have good phase coverage in CSP $YJH$, resulting in full
wavelength coverage from 3300--17500~\AA\ for each of our critical light curve
phases.  Only the older CfA targets from \citet{jha07} lack NIR data entirely,
resulting in greater uncertainties on \ebmvhost\ and \rvhost\ which we take
into account in our {analysis.}

Table \ref{tbl:snprops} lists the SNe~Ia and light curves that have passed
our selection criteria.  Light curve fit results from the \salt\ light curve
fitter \citep{guy07,guy10}, which we include for connection to previous
literature and to provide an alternative parametrization of light curve shape
for this work, \revIIb{can be found in the Online Supplementary Material.}

Unobserved ultraviolet (UV) flux bluewards of 3300~\AA\ can in principle have
a dramatic effect on inferences about \MNi\ \citep{scalzo14b}.  Photometry
in this wavelength range is rarely available for targets with $z > 0.02$,
and has been published for only two SNe~Ia in our sample,
SN~2007S and SN~2008hv.  To correct for this missing flux,
we construct a template using published photometry
from a separate sample of 79~SNe~Ia observed with the
Ultra-Violet/Optical Telescope \citep[UVOT;][]{uvot} on the \emph{Swift}
spacecraft \citep{swift}.  The UV photometry was obtained from
the \emph{Swift} Optical/Ultraviolet Supernova Archive\footnote
{http://swift.gsfc.nasa.gov/docs/swift/sne/swift\_sn.html}
\citep[SOUSA;][]{sousa}.  The reductions for the light curves are based on
that of \citet{brown09}, including subtraction of the host galaxy count rates,
and using the revised UV zeropoints and time-dependent sensitivity from
\citet{breeveld11}.  Where available, we also used public optical-wavelength
spectra from the Open Supernova Catalog \citep{guillochon16}.
Further details on the template construction are provided
in \revIIb{\S\ref{sec:uv-gp}} below.

\begin{table*}
\caption{Basic SN~Ia properties and spectroscopic subclass membership}
\begin{tabular}{lcrrccrrc}
\hline 
Name & Survey & $z_\mathrm{helio}$ & $z_\mathrm{CMB}$ & $E(B-V)_\mathrm{MW}$ & Branch & Wang & SNID & Corr.$^{a}$ \\
     &        &                    &                  & (mag)                & Type   & Type & Type &             \\
\hline 
SN~1999aa & CfA  & 0.01438 & 0.01522 & 0.040  &  SS &  91T & 99aa  & NIR      \\
SN~1999dq & CfA  & 0.01433 & 0.01356 & 0.109  &  SS &  91T & 99aa  & NIR      \\
SN~2000dk & CfA  & 0.01743 & 0.01644 & 0.069  &  CL &    N & norm  & NIR      \\
SN~2001V  & CfA  & 0.01502 & 0.01606 & 0.020  &  SS &  91T &  91T  & NIR      \\
SN~2002hu & CfA  & 0.03900 & 0.03824 & 0.045  &  SS &  91T & 99aa  & NIR      \\
SN~2004ef & \csp & 0.03100 & 0.02977 & 0.056  &  BL &   HV & norm  & NIR      \\
SN~2004eo & \csp & 0.01572 & 0.01473 & 0.108  &  CL &    N & norm  & NIR      \\
SN~2004ey & \csp & 0.01580 & 0.01463 & 0.139  &  CN &    N & norm  & NIR      \\
SN~2004gs & \csp & 0.02659 & 0.02750 & 0.031  &  CL &    N & norm  & UV+NIR   \\
SN~2004gu & \csp & 0.04579 & 0.04690 & 0.026  &  SS &  91T &  pec  & UV+NIR   \\
SN~2005M  & \csp & 0.02196 & 0.02297 & 0.031  &  SS &  91T &  91T  & $\cdots$ \\
SN~2005al & \csp & 0.01241 & 0.01329 & 0.055  & --- &  --- & norm  & NIR      \\
SN~2005el & \csp & 0.01487 & 0.01489 & 0.114  &  CN &    N & norm  & $\cdots$ \\
SN~2005eq & \csp & 0.02895 & 0.02835 & 0.074  &  SS &  91T &  91T  & NIR      \\
SN~2005eq & CfA  & 0.02895 & 0.02835 & 0.074  &  SS &  91T &  91T  & $\cdots$ \\
SN~2005hc & \csp & 0.04593 & 0.04498 & 0.033  &  CN &    N & norm  & UV+NIR   \\
SN~2005hc & CfA  & 0.04593 & 0.04498 & 0.033  &  CN &    N & norm  & UV+NIR   \\
SN~2005hj & \csp & 0.05800 & 0.05695 & 0.039  &  SS &  91T & 99aa  & UV+NIR   \\
SN~2005iq & \csp & 0.03405 & 0.03293 & 0.022  &  CN &    N & norm  & NIR      \\
SN~2005ki & \csp & 0.01917 & 0.02037 & 0.032  &  CN &    N & norm  & NIR      \\
SN~2005ls & CfA  & 0.02114 & 0.02054 & 0.093  & --- &  --- & norm  & UV+NIR   \\
SN~2006S  & CfA  & 0.03210 & 0.03296 & 0.017  &  SS &    N & 99aa  & NIR      \\
SN~2006ac & CfA  & 0.02309 & 0.02394 & 0.016  &  BL &   HV & norm  & $\cdots$ \\
SN~2006ax & \csp & 0.01671 & 0.01796 & 0.050  &  CN &    N & norm  & $\cdots$ \\
SN~2006ax & CfA  & 0.01671 & 0.01796 & 0.050  &  CN &    N & norm  & NIR      \\
SN~2006bt & \csp & 0.03217 & 0.03248 & 0.050  &  CL &    N & norm  & NIR      \\
SN~2006bt & CfA  & 0.03217 & 0.03248 & 0.050  &  CL &    N &  pec  & UV+NIR   \\
SN~2006et & \csp & 0.02217 & 0.02118 & 0.019  &  CN &    N & norm  & NIR      \\
SN~2006gt & \csp & 0.04477 & 0.04364 & 0.037  &  CL & 91bg & 91bg  & UV+NIR   \\
SN~2006gz & CfA  & 0.02370 & 0.02478 & 0.023  &  SS &  pec &  pec  & NIR      \\
SN~2006kf & \csp & 0.02127 & 0.02080 & 0.247  &  CL &    N & norm  & $\cdots$ \\
SN~2006le & CfA  & 0.01742 & 0.01729 & 0.408  &  CN &    N & norm  & UV       \\
SN~2006ob & \csp & 0.05923 & 0.05825 & 0.033  & --- &  --- & norm  & UV+NIR   \\
SN~2006ot & \csp & 0.05292 & 0.05215 & 0.018  &  BL &   HV &  pec  & UV+NIR   \\
SN~2007S  & \csp & 0.01385 & 0.01502 & 0.028  &  SS &  91T &  91T  & $\cdots$ \\
SN~2007ai & \csp & 0.03166 & 0.03199 & 0.332  &  SS &  91T &  91T  & UV+NIR   \\
SN~2007as & \csp & 0.01757 & 0.01790 & 0.142  &  BL &   HV & norm  & $\cdots$ \\
SN~2007ba & \csp & 0.03849 & 0.03906 & 0.038  &  CL & 91bg & 91bg  & UV+NIR   \\
SN~2007bd & \csp & 0.03095 & 0.03194 & 0.034  &  BL &   HV & norm  & NIR      \\
SN~2007jg & \csp & 0.03710 & 0.03658 & 0.107  &  BL &   HV & norm  & UV+NIR   \\
SN~2007nq & \csp & 0.04503 & 0.04390 & 0.035  &  BL &   HV & norm  & NIR      \\
SN~2008bc & \csp & 0.01508 & 0.01571 & 0.263  &  CN &    N & norm  & NIR      \\
SN~2008bq & \csp & 0.03395 & 0.03444 & 0.090  &  CN &    N & norm  & NIR      \\
SN~2008hv & \csp & 0.01252 & 0.01358 & 0.032  &  CN &    N & norm  & NIR      \\
SN~2008ia & \csp & 0.02198 & 0.02260 & 0.228  &  BL &    N & norm  & $\cdots$ \\
\hline 
\end{tabular}
\medskip \\
\flushleft
$^a$~Limited coverage in specific wavelength ranges:  none, NIR  , late-time $U$/$u$ (at phases $> +20$~days), or both.

\label{tbl:snprops}
\end{table*}


\section{Host galaxy extinction}
\label{sec:extinction}

To retrieve a reliable bolometric light curve, correction for extinction by
dust in the host galaxy is of paramount importance.  Rigorous and robust
estimation of \ebmvhost\ and \rvhost\ requires measurements spanning a wide
range of wavelengths, which fortunately are ensured by our selection criteria.


\subsection{Fitting multi-band light curves with \snoopy}
\label{subsec:extinction-snoopy}

Our estimates for \ebmvhost\ and \rvhost\ come from the
hierarchical Bayesian model built into the \snoopy\ light curve fitter
\citep{burns11,burns14}, which samples the full posterior
distribution of \ebmvhost\ and \rvhost\ via Markov chain Monte Carlo (MCMC).
The \citet{burns14} light curve template is parametrized by a new
light curve width parameter, \sBV, proportional to the rest-frame
time interval $\Delta t_{BV}$ between $B$-band maximum light
and the date of maximum $B-V$ color of the SN~Ia
($\sBV = 1$ for $\Delta t_{BV} = 30$~days).
This parametrization more accurately captures the morphological differences
between the NIR light curves of normal and 1991bg-like events,
compared to contemporary light curve parameters like $\Delta m_{15}$.

As a Bayesian model, the \citet{burns14} extinction model can incorporate
prior information about \ebmvhost, \rvhost, and correlations between the two.
\revIIg{For our work here, we place a ``Gaussian bin'' prior on \rvhost\
as a function of \ebmvhost,} in which the data are binned by \ebmvhost\
and an independent separate Gaussian prior on \rvhost\ is placed on SNe~Ia
within each bin \citep[see figure 14 of][]{burns14}.  Due to a numerical
instability in the \citet{fitz99} reddening law at low \rvhost, we impose
a limit $\rvhost > 0.5$.  \revIIg{An improper uniform prior was used for
\ebmvhost, so that
negative values were possible; during testing, large negative \ebmvhost\
was often} a sign of poor data quality.  Of the SNe~Ia selected for our main
analysis, only one (SN~2006kf) has a negative mean \ebmvhost\ of
$-0.03 \pm 0.02$~mag, consistent with zero extinction.

Since \snoopy\ was trained on \csp\ data, it can be used directly on
\csp\ photometry.
For CfA targets, we $S$-correct the CfA (including PAIRITEL) data to the
\csp\ natural system, using the appropriate filter transmission curves and
the \citet{hsiao07} spectral template.

To estimate systematic errors in the fit parameters introduced by the
$S$-corrections, we select a joint subset of \csp\ and CfA SNe with slightly
more permissive selection criteria than for the bolometric light curve
analysis, requiring rest-frame $BVRI$ coverage but relaxing the redshift cut
and the requirement for any data beyond day~$+35$.  This criterion provides a
larger comparison sample (15~SNe~Ia) while ensuring that coverage is similar
to the bolometric light curve sample in the range of phases needed to
constrain the multi-band light curve fit parameters.  Each SN~Ia in this
subsample satisfies the light curve quality cuts on both \csp\ and CfA
multi-band photometry, so that temporal completeness should not strongly
affect results.

\begin{figure}
\resizebox{0.5\textwidth}{!}
   {\includegraphics{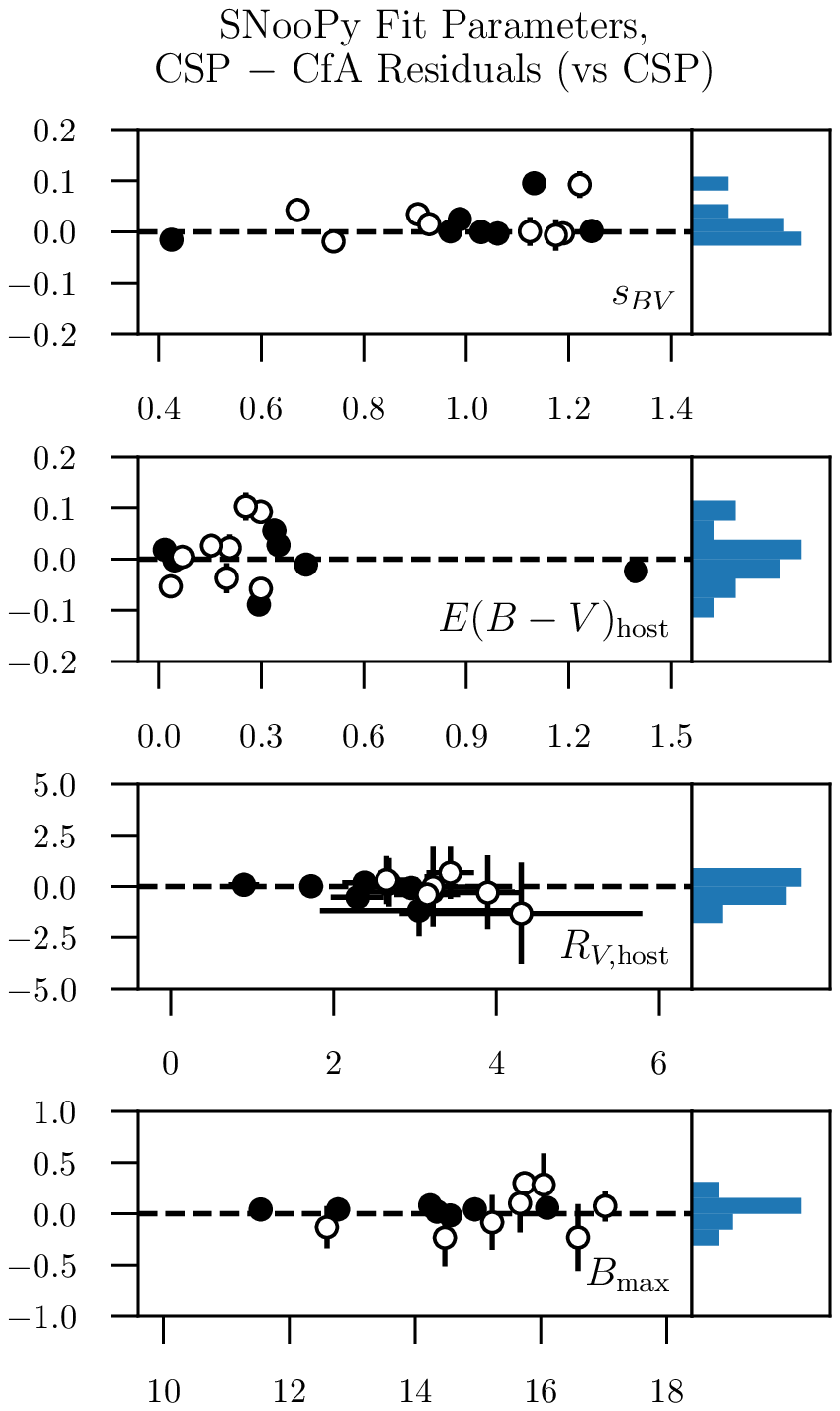}}
\caption{\small Comparison of \snoopy\ light curve fit parameters based either
on \csp\ or CfA photometry for a set of 15 SNe~Ia.  Open circles are points
for which NIR data are unavailable in one or both surveys, while filled
circles are those with NIR data from both \csp\ and PAIRITEL.  Comparisons are
shown (top to bottom) for \sBV, \ebmvhost, \rvhost, and \Bmax.}
\label{fig:snoopy-compare}
\end{figure}

Figure \ref{fig:snoopy-compare} shows comparisons of four indicative \snoopy\
outputs (\sBV, \ebmvhost, \rvhost, and \Bmax) using \csp\ data with those
obtained from $S$-corrected CfA data for the same SNe~Ia.  The correspondence
is good, though not without outliers.  Using the 68\% confidence half-width
as a robust dispersion measure gives a core dispersion of 0.03\ in \sBV\
and 0.05~mag in \ebmvhost.  The outliers tend to lack NIR and/or pre-maximum
constraints in the CfA light curve (SN~2006gj, SN~2007ai) or lie at the
extremes of \sBV\ (SN~2005ke, SN~2007S).
Estimates of \rvhost\ track each other well and are consistent within the
given uncertainties, which can be large for SNe~Ia without NIR data or
with low values of \ebmvhost.

{For purposes of our modeling, this level of accuracy in the light
curve fit parameters is adequate.  The \snoopy\ fit results for the
reddening-corrected maximum $B$ magnitude, $m_{B,\mathrm{max}}$, are
consistent within the given uncertainties for all \csp\ and CfA light curves,
so that our inferences about the peak luminosity and \nickel\ mass should be
equivalent for the two surveys.}


\subsection{Comparison with \bayessn}
\label{subsubsec:extinction-mandel}

Twenty-nine CfA SNe~Ia \revIIg{from our extinction comparison sample}
(which may not have corresponding \csp\ light curves) have
values of \ebmvhost\ and \rvhost\ derived from the \bayessn\ light curve
fitter, applied to CfA data and published in table~4 of \citet{mandel11}.
Nine of these also appear in our bolometric light curve sample.
Like the \snoopy\ ``color model'' of \citet{burns14}, \bayessn\ is a
hierarchical Bayesian light curve fitter designed to fit simultaneously
for \ebmvhost\ and \rvhost\ given optical and, when available, NIR photometry.
Although \bayessn\ can handle different prior constraints between
\ebmvhost\ and \rvhost, including the Gaussian bin prior
\citep[see ``Case 6'' of][]{mandel11}, the published extinction parameters
from \citet{mandel11} assume that \rvhost\ {varies linearly with}
\ebmvhost.  This constraint enables the fit to combine information
about \rvhost\ among all SNe~Ia, rather than only those with similar
\ebmvhost\ values {as \snoopy\ does}.
\revIIg{\bayessn\ also} uses the \citet{ccm} extinction law,
while \snoopy\ uses the \citet{fitz99} extinction law.

\begin{figure}
\resizebox{0.5\textwidth}{!}
   {\includegraphics{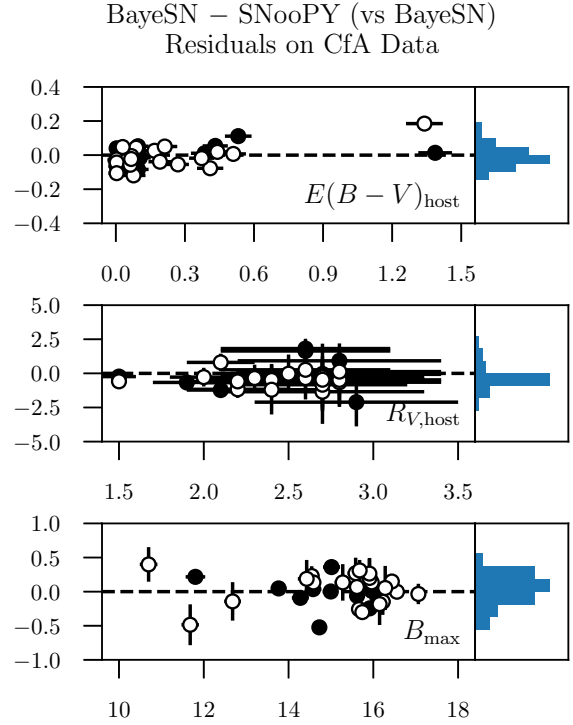}}
\caption{\small Comparison of host galaxy dust extinction parameters based
on CfA photometry for the \snoopy\ and \bayessn\ light curve fitters.
Open circles: SNe~Ia with no NIR data from PAIRITEL;
filled circles: SNe~Ia with PAIRITEL data.
Top:  \ebmvhost.  Middle:  \rvhost.  Bottom:  \Bmax.}
\label{fig:mandel-compare}
\end{figure}

The agreement \revIIg{between \snoopy\ and \bayessn}\ will be sensitive to
differences between \revIIg{their training sets and priors,
and serves as another cross-check on systematic errors in extinction.}
Figure \ref{fig:mandel-compare} shows the difference between the values of
key extinction parameters as inferred from CfA light curve data using both
\snoopy\ and \bayessn.  Results from the two fitters agree to within RMS
dispersion of 0.07~mag in \ebmvhost\ and 0.30~mag in \Bmax.  The mean
residuals in \ebmvhost\ and \Bmax\ are consistent with zero for all SNe.
The main difference between the two fitters is that the \bayessn\ results
cluster in the range $\rvhost \sim$ 2--3, with small uncertainties compared
to \snoopy; we can attribute this to \bayessn's stronger priors on allowed
values of \rvhost, and to the extinction law used.
While our fiducial results will rely on \snoopy, we run a separate inference
of \Mej\ and \MNi\ using extinction parameters from \bayessn\ wherever
results from both fitters are available and disagree significantly.


\section{Bolometric Light Curve Construction}
\label{sec:bolo}

Even with excellent broadband photometry, the bolometric light curve is not
directly observed, but is derived from simultaneous multi-wavelength data.
Despite the stringent selection criteria laid out in \S\ref{sec:obs},
sampling of light curves is still often irregular, with optical and NIR
observations being made on different telescopes at different times.
Coverage at UV or NIR wavelengths is sometimes missing entirely
and must be predicted using a plausible model of the time-evolving SED.

\revIIg{We rely on Gaussian process (GP) regression \citep{rw06}, which was
used previously by \citet{scalzo14a} in similar ways both for interpolation
in time and template correction for unobserved flux.}
We estimate the error introduced by interpolating or imputing over
missing data by deleting data points from the light curves of the targets
with the best coverage and repeating the analysis.


\subsection{Interpolation and accounting for missing data}
\label{subsec:gp}

The behavior of a GP \revIIg{as a smooth non-parametric fit} is governed by
a mean function
\revIIg{(which could be a parametric curve such as a polynomial)
and a covariance kernel that describes correlations between the residuals
of neighboring points from the mean function.
Any subset of points selected from a GP are jointly (multivariate)
Gaussian distributed, with covariance matrix given by the kernel.}
The kernel usually has hyperparameters which either are chosen to maximize
the likelihood, or marginalized over to account for all possible outcomes.
In the case of a light curve fit, for example, the hyperparameter might be
the characteristic time scale of variation in the light curve.

We use the \snoopy\ light curve fit for each SN~Ia in each band as
{the mean function for} a one-dimensional GP,
\begin{equation}
{m}_{ij} = \hat{m}_j(t_i, \Theta) + g(t_i, \Lambda),
\end{equation}
where $m_{ij}$ is the observed magnitude at time $t_i$ in band $j$,
$\hat{m}_j(t, \Theta)$ is the \snoopy\ light curve with parameters
$\Theta = (\sBV, \ebmvhost, \rvhost)$, and $g(t, \Lambda)$
is a GP fit to \revIIg{the residuals} $m_{ij} - \hat{m}_j(t_i, \Theta)$
with covariance kernel
\begin{equation}
k_\mathrm{1D}(t, t') = \exp\left[ - \frac{(t-t')^2}{\Lambda^2} \right].
\end{equation}
Random fluctuations of a given SN around the \snoopy\ fit will
\revIIg{therefore} be averaged out, while consistent deviations
\revIIg{(for example, because the target is a peculiar SN~Ia)}
will be accounted for where data are available.
At times beyond the last observation
in a given band, the model reverts smoothly to the \snoopy\ fit over the
correlation timescale $\Lambda$ of the GP (typically 1--2~weeks).  When no
data in a band are available (as in NIR bands for some targets), we simply
use the \snoopy\ predictions for that band and their associated uncertainties.

For normal SNe~Ia like those used in the \snoopy\ training set
(including many of the \csp\ objects we analyze here), systematic variations
should be minimal.  For peculiar SNe~Ia with missing data in wavelength or
phase regions that may deviate from the \snoopy\ template, we make additional
arguments about how large a deviation from the \snoopy\ template is needed
to qualitatively change our conclusions.

\begin{figure*}
\resizebox{\textwidth}{!}{\includegraphics{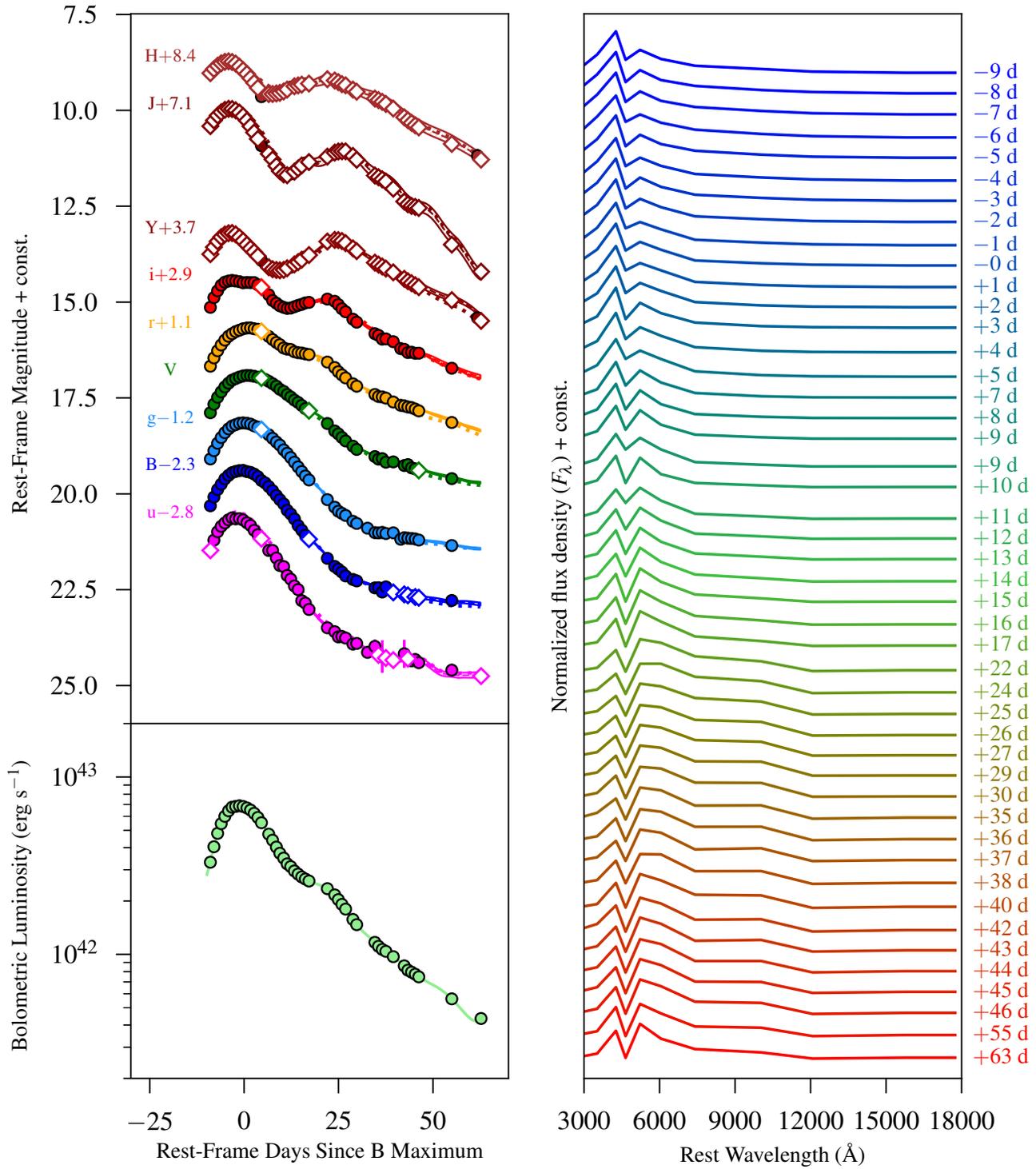}}
\caption{\small Construction of the bolometric light curve for SN~2004ef.
Top left:  broadband photometry data (filled circles) and modeled points
where data are missing (diamonds).  The \snoopy\ fit for this SN is shown as
a dotted line, and the full model taking residuals into account is shown as
a dashed line with surrounding 68\% confidence region.
Bottom left:  bolometric light curve.
Right:  coarse-resolution SED time series derived from broadband photometry.}
\label{fig:bolo-rainbow}
\end{figure*}

Figure~\ref{fig:bolo-rainbow} demonstrates the procedure on SN~2004ef,
a typical \csp\ SN~Ia.  Where data are available, the model interpolates
smoothly through them, and tracks the \snoopy\ template in bands and for
time periods where they are unavailable --- in this case, for the NIR bands
after day~$+30$.

\subsection{Correction for unobserved ultraviolet flux}
\label{subsec:uv-corr}

The \revIIg{potential variation of} the UV contribution to the bolometric
flux \revIIg{is illustrated by} two contrasting examples of well-sampled
light curves with \emph{Swift} coverage.
For the normal SN~2011fe \citep{pereira13},
flux in the range 1600--3400~\AA\ increased from the earliest
measured phases to reach a maximum of 13\% of total bolometric flux
at day~$-6$, and was close to 10\% near $B$ maximum.
For the 1991T-like SN~Ia LSQ12gdj \citep{scalzo14b},
flux in this window made up 27\% of total bolometric flux at day $-10$,
declining to 17\% by $B$ maximum and steadily thereafter.

\revIIg{Our targets do not in general have \emph{Swift} observations,
so we built a UV SED template to correct for the missing flux.  
Building a time-dependent SED template to correct for unobserved UV flux is
a challenging process, and several compromises are made; given the diversity
in UV light curve behavior, we expect to capture only the distribution of
possible UV corrections for a given SN~Ia.  We derive the correction from a
separate set of SNe~Ia observed with \emph{Swift}.}

\revIIb{A description of the full time-dependent UV correction is given in
the Online Supplementary Material (\ref{sec:uv-gp}).
Its effect is small (less than 5\% of bolometric flux) after day~$+20$.
Near maximum light, neither \sBV\ nor \vSi\ are good
predictors of the UV flux correction --- the latter potentially of interest
due to the ``NUV-red''/``NUV-blue'' subclasses posited by \citet{milne13}.
Its main influence is therefore as a systematic error on bolometric flux at
maximum light, which for purposes of deriving \nickel\ mass
can be treated as random.}

\begin{figure}
\resizebox{0.47\textwidth}{!}{\includegraphics{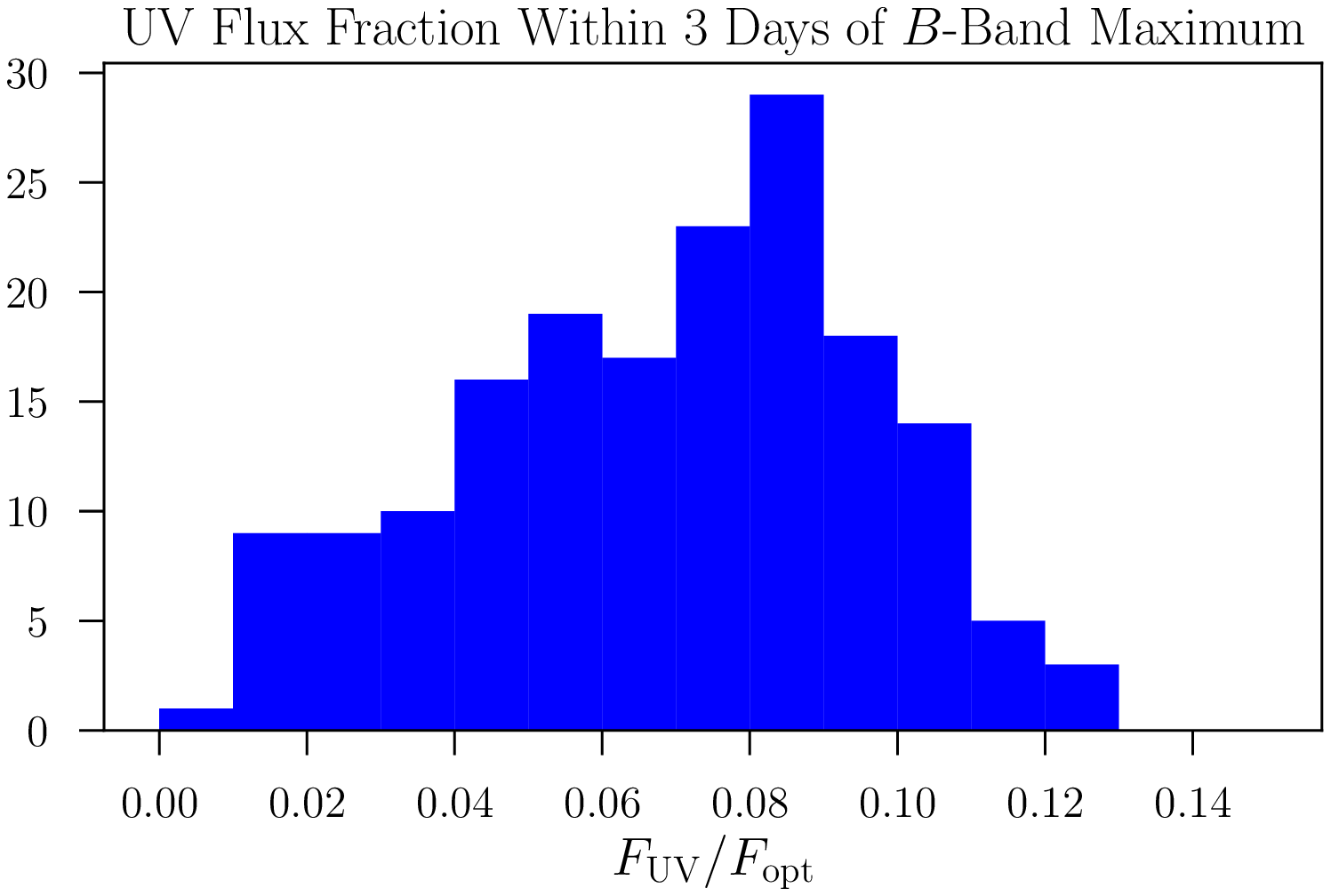}}
\caption{\small Ratio of UV (1600--3300~\AA) to optical (3300--8500~\AA)
flux within 3 days of $B$-band maximum light.}
\label{fig:uv-frac-atmax}
\end{figure}

Figure~\ref{fig:uv-frac-atmax} shows the distribution of UV flux fraction
within 3~days of the date of maximum light from the \snoopy\ fit.
The distribution peaks around 0.08
(with mean 0.07 and standard deviation 0.03), but is skewed towards lower
values.  The assumption by \citet{scalzo14a} of a uniform distribution
between 0.0 and 0.1,
\revIIb{contributing a systematic error of about 3\% to \MNi,}
is shown to be slightly biased in the mean, but not catastrophically wrong.


\subsection{Building the bolometric light curves}
\label{subsec:bolo}

For each target, given a suite of light curves with quasi-simultaneous
measurements in a range of bands, we build the bolometric light curve by
the procedure used in \citet{scalzo14b}.
We briefly summarize the process here.

Photometric measurements taken at similar times across different bands
are grouped into single multi-band measurements, each representing at least
four measurements within a 0.2-day window.  The GP model is evaluated in
each band to interpolate missing values at the mean date
of matched observations.  Missing values are interpolated
only for phases before day~$+70$, since {the \citet{hsiao07}
spectrophotometric time series template, used by \snoopy\ for bandpass
corrections, ends at} this phase.

For each multi-band measurement, a piecewise linear broad-band SED is
constructed in the observed frame \revIIg{using the ``best-fit SED''
method of \citet{brown16}}, then de-redshifted to the rest frame.
Full $K$+$S$-corrections are not computed, since flux from one band will
be shifted into neighboring bands and will still be captured in the total
bolometric flux.

The resulting optical-wavelength SED is corrected for Milky Way dust
extinction using the \citet{sf11} recalibration of the \citet{sfd98}
dust maps.  \revIIg{The correction for host galaxy dust extinction is often
much larger and more uncertain} than the Milky Way extinction; rather than
applying a single mean correction, corrected SED time series are generated
\revIIg{to cover a grid of values}
of \rvhost\ from 0.0--10.0 at 0.2~mag/mag intervals,
and of \ebmvhost\ from 0.00--0.50 at 0.02~mag intervals.  Unextinguished UV
flux densities are predicted from the GP template described in
\S\ref{sec:uv-gp}, normalized to the flux density point corresponding
to $B$-band, and \revIIg{joined} to the optical-wavelength SED.
For each value of \ebmvhost\ and \rvhost,
\revIIg{the resulting UV-optical SED is integrated}
from 1600--17500~\AA\ to obtain the bolometric flux as a function of time.

As a cross-check, ``leave-data-out'' tests \revIIg{are performed},
where NIR and late-time $U$/$u$ points \revIIg{are removed}
from our best-covered targets, \revIIg{and the light curves are reconstructed
and compared to the original versions}
at all points between day~$+20$ and day~$+70$ with respect to the date of
maximum light from the \snoopy\ fit.
For our nine targets with 3300--20000~\AA\ coverage or equivalent at allo
critical epochs, the bolometric flux is unchanged to less than 2\%~RMS.
The residual distribution broadens to 3\% RMS when the entire sample is
considered.  Some peculiar SNe~Ia, such as SN~2006ot, show greater deviations
of up to 10\% when the template is used, demonstrating the importance of
good temporal and wavelength coverage for peculiar events.

Our bolometric light curves can be found \revIIb{in ASCII format
in the Online Supplementary Material}.  For the reader's convenience in
computing light curves under different estimates of \ebmvhost, \rvhost,
and distance without exhaustively tabulating all values, we provide the
observer-frame, unreddened ($\ebmvhost = 0.0$) time-dependent bolometric
flux, $f_\mathrm{bol,0}(t)$, and the coefficients of a fitting formula
of the form
\begin{eqnarray}
\log_{10} f_\mathrm{bol}(t)
    &\!\! = \!\! & \log_{10} f_\mathrm{bol,0}(t) \nonumber \\
    & + & a_C(t) \times \ebmvhost \nonumber \\
    & + & a_{RC}(t) \times \rvhost \times \ebmvhost \nonumber \\
    & + & a_{RCC}(t) \times \rvhost \times (\ebmvhost)^2,
\end{eqnarray}
converting to isotropic luminosity via the luminosity distance $d_L$,
\begin{equation}
L_\mathrm{bol}(t) = 4\pi d_L^2 f_\mathrm{bol}(t).
\end{equation}
The expansion provides results with an absolute deviation limited to 0.01~dex
(2.3\%) worst-case for $\rvhost < 5$ (suitable for all SNe in this paper).
For $\rvhost < 5$ and $\ebmvhost < 0.3$~mag, suitable for all except our two
most reddened SNe~Ia, the worst-case deviation drops to 0.004~dex (0.9\%).


\section{Modeling Procedure}
\label{sec:bolomass}

For bolometric light curve modeling, we use the bolometric light curve suite
\bolomass\footnote{\texttt{github.com/rscalzo/pyBoloSN}}
\revIIg{\citep{scalzo14a,scalzo14b}}.  \bolomass\ is a
freely-available Python-based toolkit for Bayesian probabilistic inference
of {global SN~Ia explosion properties from} bolometric light curves.
\revIIb{The inference proceeds by sampling \Mej, \MNi, and other global
properties of the white dwarf progenitor(s), explosion mechanism, and
observational circumstances of each SN~Ia, collectively denoted by $\theta$.
A semi-analytic forward model predicts the bolometric light curve
$\Lbol{fwd}(\theta; t_i)$,
which is then compared to the data $\Lbol{obs}(t_i)$ using Bayes's rule:
\begin{equation}
P(\theta | \Lbol{obs}) =
    \frac{P(\Lbol{obs} | \theta) P(\theta)}{P(\Lbol{obs})}.
\end{equation}
The log likelihood $\log P(\Lbol{obs} | \theta)$ is Gaussian,
with mean given the forward model $L(\theta; t_i)$ and variance given by
the independent observational uncertainties $\sigma_{L,i}$.
The prior $P(\theta)$ encodes pre-existing knowledge about the parameters
$\theta$, which include uncertainties on ``nuisance'' parameters
such as host galaxy reddening and distance, as well as physical constraints
(such as conservation of energy, nucleosynthesis, or radiation transfer).
The evidence
$P(\Lbol{obs}) = \int P(\Lbol{obs} | \theta) P(\theta) \, d\theta$
is a normalizing constant that can be ignored as long as all models of
interest lie within the parameter space spanned by $\theta$, which can be
sampled by MCMC.}

The model system is a spherically symmetric, homologously expanding density
distribution of ejecta.  The composition is parametrized by four broad
categories of elements:  stable iron, \nickel, intermediate-mass elements,
and unburned carbon/oxygen, lying in concentric spherical shells of density
decreasing with increasing velocity.
Although the model contains no hydrodynamics, it parametrizes turbulent
mixing between shells by a mixing length \aNi\
(in mass fraction coordinates) over which the composition changes smoothly
from one shell to another, as in \citet{kasen06}.
Physics such as white dwarf rotation and neutronization at high densities
is also parametrized \citep{yl05,krueger12}.  Full numerical
simulations of radiation transfer in SN~Ia atmospheres with similar
features, such as \code{cmfgen} \citep{cmfgen2,cmfgen1}, have been quite
successful in describing the overall features of SN~Ia
photometric and spectroscopic evolution.  \revIIb{Where possible,
we minimize model dependence by prioritizing ``consensus'' constraints,
based on conservation laws or on correlations seen in multiple codes.}

In our approach, the peak luminosity contributes the most direct constraint
on \MNi, as in ``Arnett's rule'' \citep[after][]{arnett82}.  Following other
authors \citep[e.g.][]{branch92,stritz06,howell06,howell09}, we include as a
nuisance parameter the ratio
$\alpha = L_\mathrm{max,bol}/L_\mathrm{radio} \sim 1$ of
bolometric luminosity to instantaneous energy release by radioactive decay,
accounting for opacity variation not captured by the \citet{arnett82} model.
We also take differences in rise times into account, although the full
pre-explosion rise is not in general observed in our
{data; a prior on rise time is}
incorporated through its dependence on decline rate \citep{ganesh11}.
At late times, we approximate
energy transport of \cobalt-decay gamma rays
{in the Compton-thin regime} as a constant
$\kgamma = 0.025$~cm$^2$~g$^{-1}$ \citep{swartz95},
and calculate the {mean gamma-ray} optical depth based on the radial
distribution of \nickel\ \citep{jeffery99}.
Our model is sensitive to the \nickel\ distribution to
at least this extent, thus sharing some features of the semi-analytic models
of \citet{pe00a}, although we do not try to predict or interpret the detailed
bolometric light curve pre-maximum, or between $B$ maximum and day~$+40$.

\revIIb{
The capacity of the Bayesian paradigm to incorporate \emph{informative}
prior information is a double-edged sword:  informative priors can reduce
the posterior uncertainty, but the results may also be sensitive to the
prior used.  Nuisance parameters are major contributors to the final
uncertainty in our inference and so our assumptions about them matter.
We therefore run several different scenarios corresponding to different
informative prior assumptions about the physics of radiation transfer in the
expanding supernova atmosphere:
\begin{enumerate}
\item Our fiducial analysis uses the ``Run~F'' priors
      of \citet{scalzo14a},
      which assume $\alpha = 1$ and no dense core of iron-peak elements,
      and were validated in that work through a blind trial against
      a suite of 3-D numerical explosion models.  These also produce
      predictions close to the median \Mej\ for the eight different priors
      explored in \citet{scalzo14a}.
\item Two additional runs replace the assumption of $\alpha = 1$ with
      empirical priors that emulate ensemble-average correlations
      between $\alpha$, \nickel\ content, and white dwarf central density,
      as estimated from the model grids of \citet{hk96}
      and \citet{blondin13,blondin17}.
\item Finally, two additional runs under the fiducial priors modify the
      light curve, to evaluate the impact of potential missing flux at
      mid-infrared (MIR) wavelengths.  These corrections are inspired by
      numerical simulations of Chandrasekhar-mass models, and so they may
      only be applicable conditional on other global parameters
      (e.g., $\Mej \sim \Mch$), but are applied uniformly without regard
      to other parameter covariances.  Our expectation based on
      prior experience is that an increase in late-time flux will
      increase the inferred mass, especially of sub-Chandrasekhar-mass
      candidates.  We consider MIR contributions near day~$+60$ after
      bolometric maximum of either 10\% (estimated for normal SNe~Ia)
      or 25\% (estimated for 1991bg-like SNe~Ia).
\end{enumerate}
The Online Supplementary Material (\ref{sec:priors}) provides detailed
justifications of each of these
choices of priors and described how they were implemented.}

\revIIg{In addition to these physical priors, we treat \ebmvhost, \rvhost,
and $d_L$ as nuisance parameters increasing the uncertainty on \Mej\ and \MNi.
Our analysis calculates the luminosity distance
assuming a flat $\Lambda$CDM model ($\Omega_M = 0.3$,
$\Omega_\Lambda = 0.7$, $H_0 = 70$~\kms~Mpc$^{-1}$),
and a 300~\kms\ systematic uncertainty on the SN~Ia redshift
from random peculiar velocities.}

\revIIb{We use the \emcee\ package \citep{emcee} to perform MCMC sampling
over the model parameters.  \citet{scalzo14a} note that the
posterior distribution $P(\Mej, \MNi | \mathrm{data})$ for any given SN~Ia
may be bimodal, with one mode at $\Mej < \Mch$ and one with $\Mej \sim \Mch$.
Accordingly, we follow \citet{scalzo14a} in using \emcee's parallel-tempered
MCMC sampler for our work in order to explore both modes thoroughly.}

\begin{figure*}
\resizebox{\textwidth}{!}{\includegraphics{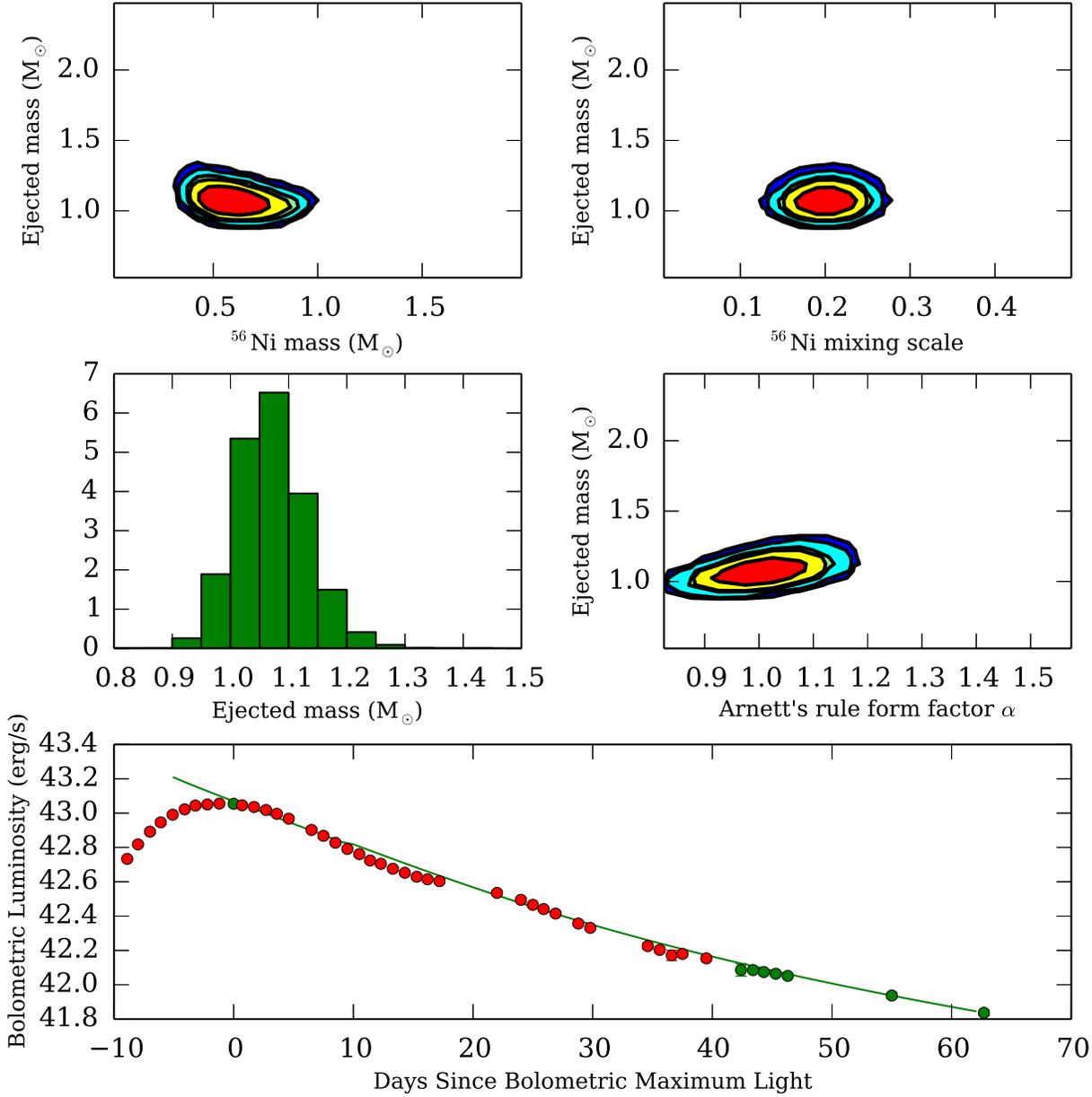}}
\caption{\small Confidence regions of progenitor properties for SN~2004ef.
Contour plots represent projections of the full joint distribution of all
parameters into the plan spanned by the panel axes.  Contours bound regions
of constant probability density.  Colored regions, moving radially outwards,
are 68\% (red), 90\%, 95\%, 99\%, and 99.7\% confidence level (blue).
Bottom:  bolometric light curve of SN~2004ef.  Green circles show which data
points are included in the fit, in regions where the
{approximations made by the} semi-analytic energy deposition model
{are expected to be valid; red points are excluded from the fit.}
{The green curve shows} the maximum \emph{a posteriori}
rate of radioactive energy {deposition in the ejecta}
(multiplied by the form factor $\alpha$,
which is in this case close to 1, near bolometric maximum).}
\label{fig:bolo-contours}
\end{figure*}

\revIIg{Figure~\ref{fig:bolo-contours} shows a summary of the MCMC fit for
SN~2004ef under our fiducial priors.  The semi-analytic expression
used for the radioactive energy deposition provides an excellent description
of the bolometric light curve after day~$+40$.  While $\alpha$ is
permitted to vary, a value near 1.0 suffices to describe the data well.}

\section{Modeling Results}
\label{sec:results}

\revIIb{Tables of the global explosion parameters inferred from our modeling,
in ASCII format, can be found in the Online Supplementary Material.}
As in \citet{scalzo14a}, we find a range
of 0.9--1.5~\Msol\ for \Mej\ and 0.4--1.1~\Msol\ for \MNi, with the massive,
\nickel-rich end dominated by slow-declining SNe~Ia
(including {SNe~Ia spectroscopically similar to SN~1991T})
and the less-massive, \nickel-poor end by fast-declining SNe~Ia
(including {SNe~Ia spectroscopically similar to SN~1991bg}).
Where light curves from both surveys are available, we analyze them
independently, and find {agreement within the uncertainty estimates
given by our modeling.}

\subsection{\revIIg{Number of non-standard explosions and prior sensitivity}}

\revIIg{In the fiducial analysis, the inferences for 19 out of 41 SNe~Ia show}
$\Mej < 1.4$~\Msol with $> 95\%$ \revIIb{probability}
(almost all of these at $> 99$\% \revIIb{probability}).
For 10 other events, $\Mej > 1.4$~\Msol at $> 95\%$ \revIIb{probability},
although for some of these (notably SN~2001V and SN~2005ls), the formal
\revIIb{credible} intervals \revIIg{are more sensitive} to assumptions about
host galaxy dust extinction than for the sub-Chandrasekhar-mass candidates.
In the cases of SN~2006bt and SN~2006ot, the host galaxy extinction parameters
from \snoopy\ are believed to be unreliable, although \sBV\ may still be used
as a description of the light curve shape.  We re-run the fits for all of
these targets under different reddening assumptions to test the robustness of
our conclusions.  We give more detailed comments on analysis assumptions and
cross-checks for individual SNe~Ia in
\revIIb{the Online Supplementary Material (\ref{sec:sncomments}).}

\revIIg{The \citet{hk96} prior on $\alpha$ tends to increase the
median posterior value of $\Mej$ by up to 0.1~\Msol, and incurs larger
uncertainties on both \Mej\ and \MNi.  As a result, fewer individual SNe~Ia
are identified as non-Chandrasekhar mass at high probability than in
the fiducial analysis.
In contrast, the prior trained on the \citet{blondin17} models produces
results indistinguishable from our fiducial analysis for most SNe,
perhaps because it does not differ strongly from $\alpha = 1.0$ except
for events with low $\MNi/\Mej$, such as SN~2006ot.}

\revIIg{We expect the approximate MIR flux correction to result in an
increase in inferred \Mej, since it mainly modifies the late-time light curve.
This is indeed what happens, with the fractional increase in \Mej\
being comparable to the fractional flux increase at day $+60$.
Most (14/19) SNe~Ia inferred to be sub-Chandrasekhar in our fiducial analysis
remain sub-Chandrasekhar {under the 10\% correction}.
Even under the more extreme {25\% correction}, which should apply only to
SNe~Ia with luminosities and decline rates typical of the 1991bg-like
subclass, three of our candidates remain sub-Chandrasekhar at a formal
probability greater than 99\%: SN~2000dk, SN~2006gt, and SN~2006kf.}

The probability of any single SN~Ia having $\Mej \neq \Mch$
may depend sensitively on the details of the reconstruction for that SN~Ia,
including the priors used on the approximated explosion physics.
However, we expect the total number of SNe~Ia in the sample falling in
different mass brackets to be more robust, since errors in \Mej\ for
different SNe~Ia will be independent provided that our modeling has captured
covariances between \Mej\ and other variables.  We assess this by drawing
1,000 simulated datasets, each containing a posterior draw of \Mej\ for
each SN~Ia, under each of the three priors.
Following \citet{scalzo14c}, for each dataset
\revIIg{the simulated \Mej\ values are binned} in three bins corresponding
roughly to different explosion scenarios:
``sub-Chandra'' (below 1.35~\Msol),
``Chandra-mass'' (1.35--1.5~\Msol, allowing for rapid rotation), and
``super-Chandra'' (above 1.5~\Msol).
The results are listed in Table~\ref{tbl:NMCh}; there is little
variation in the \emph{total} predicted number of SNe~Ia in each mass bin.

\begin{table}
\caption{Number of expected SNe~Ia in each mass bin}
\begin{tabular}{lrrr}
\hline 
Prior                  & $N(<\Mch)^a$ & $N(\Mch)^b$ & $N(>\Mch)^c$ \\
\hline 
Run~F ($\alpha = 1.0$) & $22.6 \pm 1.6$ & $ 8.4 \pm 2.2$ & $11.9 \pm 1.8$ \\
HK96 $\alpha$ Prior    & $20.3 \pm 2.0$ & $ 8.4 \pm 2.3$ & $14.3 \pm 2.1$ \\
B17 $\alpha$ Prior     & $22.0 \pm 1.4$ & $ 8.5 \pm 2.1$ & $12.4 \pm 1.8$ \\
\hline 
\end{tabular}
\medskip \\
\flushleft
\emph{Notes.} Uncertainties reflect the standard deviation of counts of
    simulated SNe~Ia within each given mass range.
$^a$~Number of ``sub-Chandra'' SNe~Ia with $\Mej < 1.35~\Msol$. \\
$^b$~Number of ``Chandra-mass'' SNe~Ia with $1.35~\Msol < \Mej < 1.5~\Msol$. \\
$^c$~Number of ``super-Chandra'' SNe~Ia with $\Mej > 1.5~\Msol$. \\

\label{tbl:NMCh}
\end{table}


\subsection{\Mej\ and \MNi\ vs. multi-band light curve width parameters}
\label{subsec:modeling-sBV}

\begin{figure*}
\resizebox{\textwidth}{!}{\includegraphics{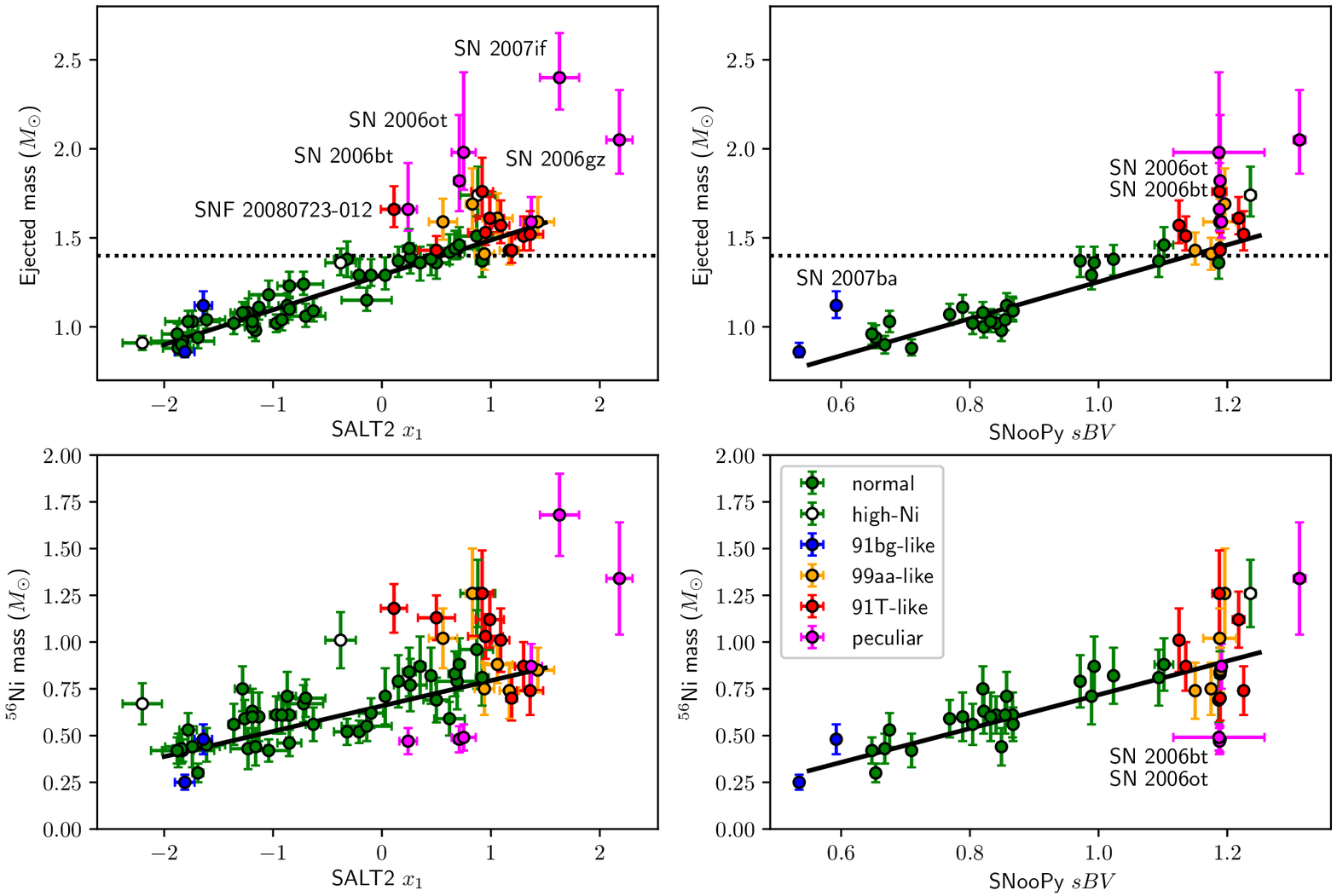}}
\caption{\small Ejected mass \Mej\ {(top)}
and \nickel\ mass \MNi\ {(bottom)}
plotted against \salt\ $x_1$ {(left)}
and \snoopy\ \sBV\ {(right)}.
Colors show different spectroscopic subtypes as revealed by \snid:
1991T-like (red), 1999aa-like (orange), normal (green);
1991bg-like (blue); and other peculiar (magenta).
{The open points represent spectroscopically normal SNe~Ia
with anomalously high inferred \MNi, which have been excluded from
the best-fit linear trend(s).}}
\label{fig:mejvsBV}
\end{figure*}

Figure~\ref{fig:mejvsBV} shows correlations of \Mej\ and \MNi\ with
\salt\ $x_1$ and \snoopy\ \sBV.  The top two panels also include the
\revIIg{fiducial}
values of \Mej\ and \MNi\ as functions of $x_1$ for the SNfactory sample of
\citet{scalzo14a}.  For normal SNe~Ia, all of these correlations can be
described by fits to simple linear relations.

\revIIg{We update our fitting formulae} using our fiducial reconstructions
for all spectroscopically normal \revIIg{SNe~Ia}, excluding high-\nickel\
outliers: the SNfactory light curves of SN~2005el and SNF~20070701-005,
\revIIg{which were previously excluded in \citet{scalzo14a},}
and the CfA light curve of SN~2005ls, which
has uncertain reddening and at-maximum spectroscopic behavior.
Fitting the remaining 42 data points yields
\begin{equation}
\Mej = (1.291 \pm 0.014) + (0.196 \pm 0.011) \, x_1
\end{equation}
with $\chi^2/\nu = 28.6/43 = 0.72$, and
\begin{equation}
\MNi = (0.659 \pm 0.023) + (0.136 \pm 0.019) \, x_1
\end{equation}
with $\chi^2/\nu = 63.3/43 = 1.06$.
The parametrizations of other commonly used light curve fitters
(\code{SiFTO} $s$, MLCS2k2 $\Delta$) can be smoothly transformed to and from
$x_1$, enabling these linear relationships to be transferred readily into
analogous results for other light curve fitters.

Since \snoopy\ \sBV\ does not map uniquely or smoothly to and from $x_1$,
we fit new relations here.  The best-fit linear relations
for {\Mej\ and} \MNi\ vs. \sBV, using only \csp\ + CfA data for
which the light curve fits are available, is {
\begin{equation}
\Mej = (1.253 \pm 0.021) + (1.036 \pm 0.095) \times (\sBV-1)
\end{equation}
($\chi^2/\nu = 27.0/24 = 1.12$), and
\begin{equation}
\MNi = (0.718 \pm 0.027) + (0.903 \pm 0.108) \times (\sBV-1)
\end{equation}
($\chi^2/\nu = 18.4/24 = 0.77$).
The targets are not distributed uniformly along the \sBV\ axis, and so it
remains unclear whether the underlying true dependence of \Mej\ on \sBV\
might be more complex; for example, a step-function transition from
$\Mej = 1.00$~\Msol\ to $\Mej = 1.39$~\Msol\ around $\sBV = 0.9$
describes the data equally well ($\chi^2/\nu = 25.8/23 = 1.12$).  Additional
data could in the future clarify the functional form, and determine whether
the photometric regularities picked up by \sBV\ are reflected in the inferred
global physical parameters of the explosion.}


\subsection{\Mej\ vs. SN spectroscopic properties}
\label{subsec:modeling-vSi}

\begin{figure}
\resizebox{0.47\textwidth}{!}{\includegraphics{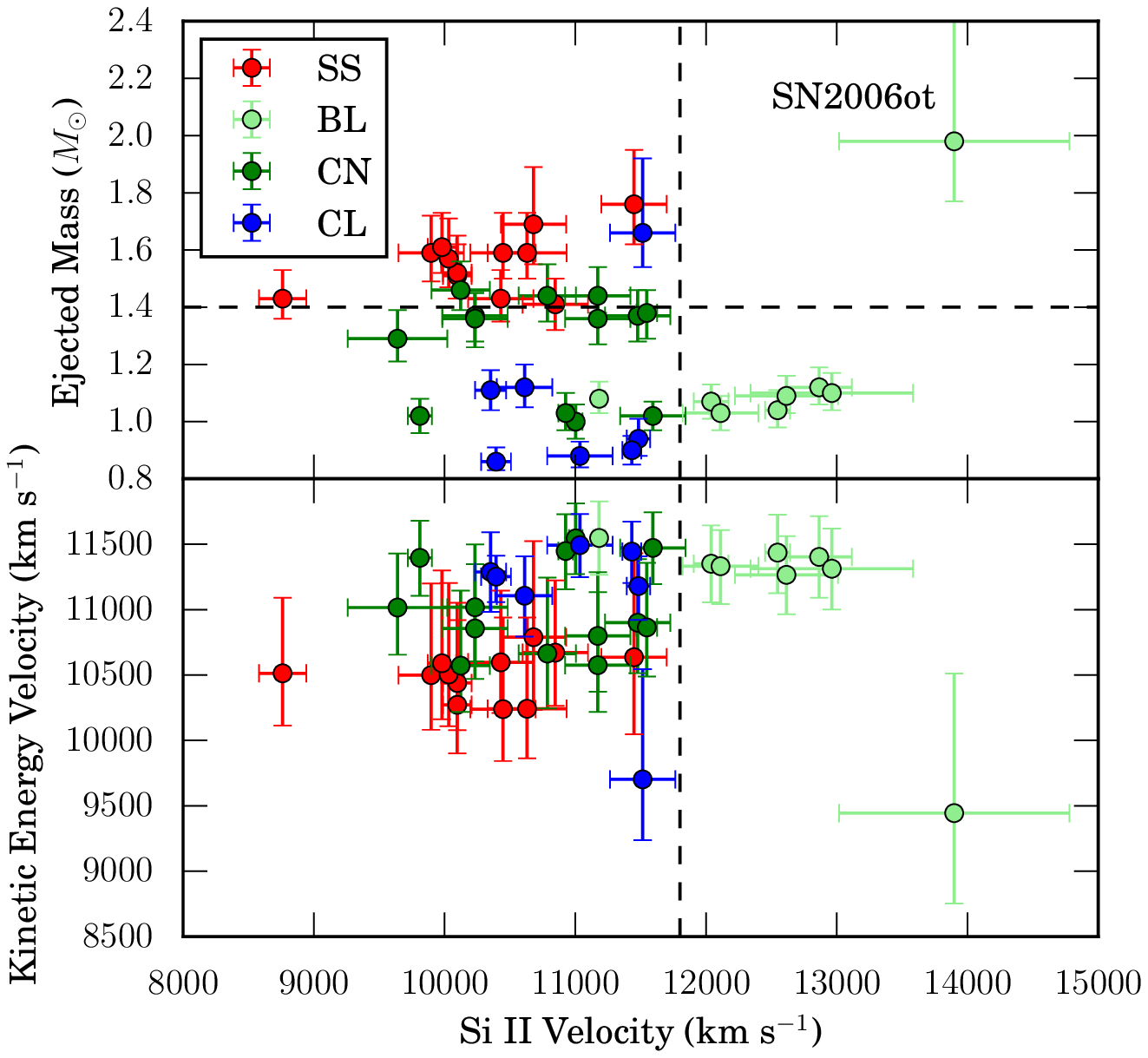}}
\caption{\small Ejected mass (top) and inferred kinetic energy velocity
(bottom) vs. \ion{Si}{2} velocity \vSi\ at $B$-band maximum light.
Colors represent \protect\citet{branch06} spectroscopic subtype:
shallow-silicon (SS; red), broad-line (BL; light green),
core-normal (CN; dark green), and cool-photosphere (CL; blue).
The dashed horizontal line at top marks $\Mej = \Mch$.
The dashed vertical line marks the boundary between the \citet{wang09}
``normal'' and ``high-velocity'' subtypes.}
\label{fig:mejvvsi}
\end{figure}

Figure~\ref{fig:mejvvsi} shows the sample broken down by \citet{branch06}
spectroscopic subtype and by \vSi, \revIIg{which determines membership in
the \citet{wang09} ``normal'' (N) and ``high-velocity'' (HV) subtypes.
The Branch type is a good predictor of
which quadrant of the top panel each SN falls into.}
Shallow-silicon (SS) events are consistently Wang-N events with
\revIIg{inferred $\Mej \geq \Mch$},
while core-normals (CN) and cool-photosphere (CL) events
are consistently Wang-N events with \revIIg{$\Mej \leq \Mch$}.
Broad-line (BL) events map well to the Wang-HV subclass, and cluster
within a narrow range of sub-Chandrasekhar masses, with the
exception of SN~2006ot.

\revIIg{The bottom panel of Figure~\ref{fig:mejvvsi} compares the inferred
kinetic energy velocity \vKE\ to the measured \vSi,}
which is often used as a proxy for kinetic energy in the literature
\citep[e.g.,][]{fk11}.  \revIIg{Little correlation is seen between the two}
(Pearson rank $r = 0.09$); \revIIg{apart from
the split between Wang-N and Wang-HV subclasses,
\vSi\ seems to be a better predictor of the density and ionization state of
the outer layers of ejecta above the \ion{Si}{2} layer than the velocity of
the bulk ejecta underneath it.}


\subsection{\Mej\ vs. host galaxy properties}
\label{subsec:modeling-Mhost}

\begin{figure}
\resizebox{0.47\textwidth}{!}{\includegraphics{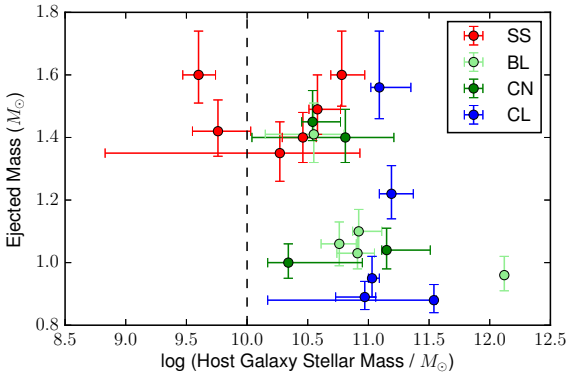}}
\caption{\small Ejected mass vs. host galaxy stellar mass.  Colors of markers
indicate the same Branch types as in Figure~\ref{fig:mejvvsi} above.
The dashed vertical line at $M_\mathrm{host} = 10^{10}~\Msol$ marks the
division between ``low-mass'' and ``high-mass'' galaxies used in contemporary
cosmology analyses \protect\citep[e.g.,][]{betoule14}.}
\label{fig:mejvmhost}
\end{figure}

In Figure~\ref{fig:mejvmhost} we plot the ejected masses for our sample
against the stellar masses of their host galaxies.  We see that SNe~Ia with
low-mass progenitors appear in high-mass galaxies, while low-mass galaxies
tend to produce SNe~Ia with more massive progenitors.  This is the expected
result given the correlation of ejected mass with stretch and the
well-established correlation of stretch with host galaxy mass
\citep[e.g.][]{branchvdb93, hamuy96, hamuy00, howell09}.

Interestingly, Figure~\ref{fig:mejvmhost} may indicate that SNe~Ia
transition from being predominantly Chandrasekhar-mass to predominantly
sub-Chandrasekhar-mass at a host galaxy mass scale of about
$\log(M/M_\odot) \sim 10.5$.
\citet{cwz14} showed that the mean ages of SNe~Ia also undergoes a
sharp transition around the same galaxy mass scale.  This would indicate that
the ejected mass (i.e. progenitor mass) and thus stretch may be driven by the
age of the SN~Ia progenitor system.  This has been suggested in the past
\citep[e.g.,][]{hamuy96, howell01}, with proposed explanations such as the
age dependence of a white dwarf's carbon-to-oxygen ratio \citep{umeda99}.


\subsection{Trends with bolometric light curve morphology}

\newcommand{\thp}{\ensuremath{t_{+1/2}}}
\newcommand{\thm}{\ensuremath{t_{-1/2}}}

We now turn to \revIIg{correlations} between \Mej, \MNi,
and morphological properties of bolometric light curves:
\begin{enumerate}
\item the bolometric luminosity \Lbol{max};
\item the late-time luminosity \Lbol{40};
\item the bolometric light curve decline rate \dmbol{15}, measured as the
      difference between the magnitude at bolometric maximum and 15 days
      after bolometric maximum;
\item the late-time decline rate \dmbol{40}, defined similarly with respect
      to the luminosity 40~days after bolometric maximum;
\item \thp, the time in days for the bolometric luminosity to decline from
      maximum to one-half maximum luminosity;
\item \thm, the time in days for the bolometric luminosity to rise from
      one-half maximum to maximum luminosity
      (where light curve completeness at early phases permits).
\end{enumerate}
\revIIg{These properties are measured} by evaluating the GP-based residual
model directly at the required epochs, using it to interpolate smoothly
in time.  \revIIg{Marginalizing} the model over \ebmvhost\ and \rvhost\
\revIIg{shows that} the variation is {at or beneath} the systematic error
floor of 3\% established by the leave-data-out comparisons.
\revIIb{The Online Supplementary Material} \revIIg{contains these measurements
for bolometric light curves in the present work, 
for the published SNfactory bolometric light curves from \citet{scalzo14a}}
\revIIb{(though with a larger systematic error of 0.1~mag on \dmbol{15}\ due
to the choice of parametrization for the NIR corrections)},
\revIIg{and of synthetic
bolometric light curves from a suite of 3-D explosion models under
various scenarios spanning a range of \Mej\ and \MNi\
\citep{kromer10,pakmor12,ruiter13,seitenzahl13a}.
In total, 63 real SNe~Ia and 8 models are shown.}

\begin{figure*}
\resizebox{\textwidth}{!}{\includegraphics{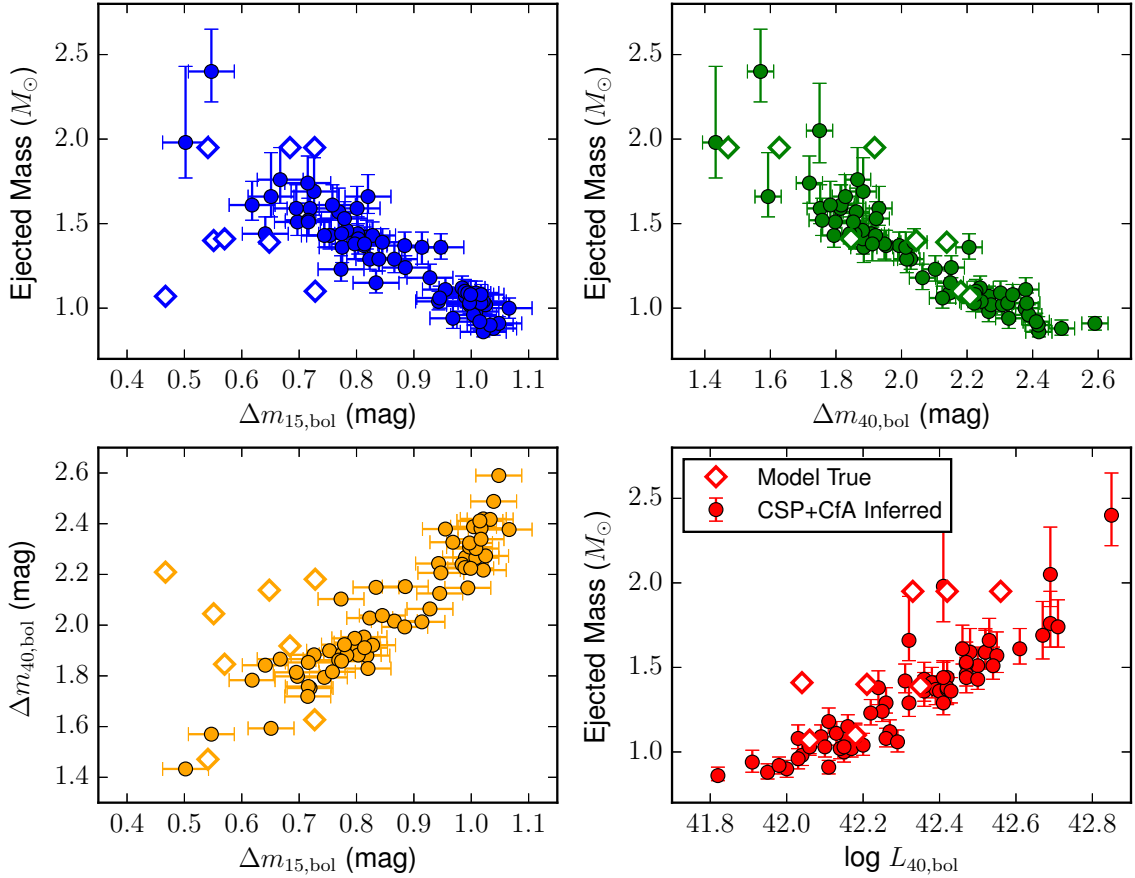}}
\caption{\small Correlations of ejected mass \Mej\ with bolometric light
curve properties:
$\Delta m_\mathrm{15,bol}$ (top left), 
$\Delta m_\mathrm{40,bol}$ (top right), 
and $L_\mathrm{40,bol}$ (bottom right).
Bottom left:  $\Delta m_\mathrm{40,bol}$ vs. $\Delta m_\mathrm{15,bol}$.
Filled circles:  SN~Ia data.  Open diamonds:  explosion models.}
\label{fig:lcpropsvmej}
\end{figure*}

\begin{figure*}
\resizebox{\textwidth}{!}{\includegraphics{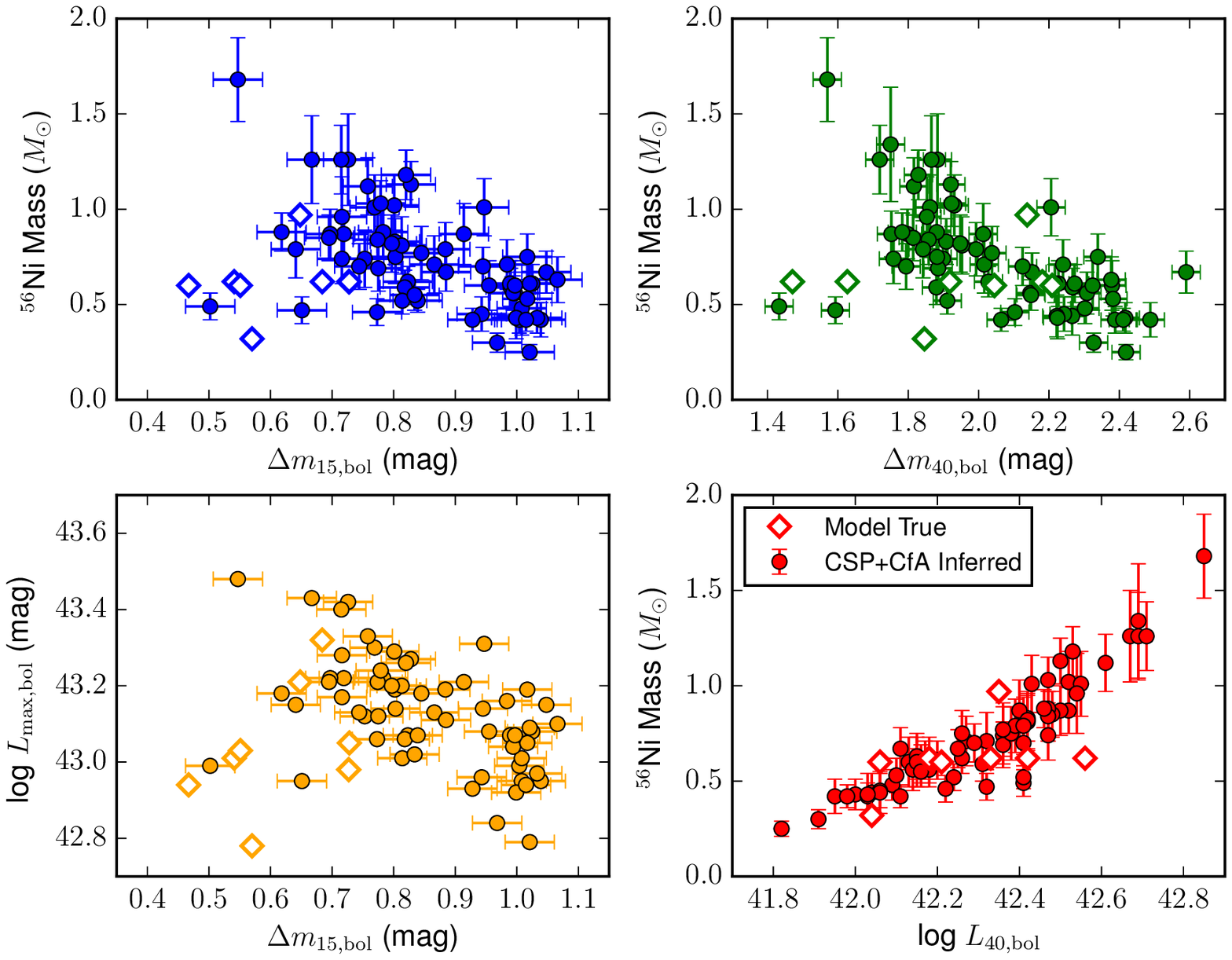}}
\caption{\small Correlations of \nickel\ mass \MNi\ with bolometric light
curve properties:
$\Delta m_\mathrm{15,bol}$ (top left), 
$\Delta m_\mathrm{40,bol}$ (top right), 
and $L_\mathrm{40,bol}$ (bottom right).
Bottom left:  $L_\mathrm{max,bol}$ vs. $\Delta m_\mathrm{15,bol}$.
Filled circles:  SN~Ia data.  Open diamonds:  explosion models.}
\label{fig:lcpropsvmni}
\end{figure*}

\revIIg{Figure~\ref{fig:lcpropsvmej} shows that} \dmbol{40}\ is an excellent
predictor of \Mej, as \revIIg{suggested by} figure~6 of \citet{scalzo14a}
but not made explicit.  The extremely high-mass SNe~Ia
(SN~2006bt, SN~2006ot, SN~2006gz and SN~2007if) all have $\dmbol{40} < 1.6$,
\revIIg{separated} from the normal and 1991T-like SNe~Ia,
\revIIg{which all have $\dmbol{40} > 1.7$}.
The explosion models lie along roughly the same locus as the real SNe~Ia,
although the three lines of sight for the asymmetric violent merger model
11+09 \citep{pakmor12} show more variation than the Chandrasekhar-mass and
sub-Chandrasekhar-mass models, which are closer to being spherically symmetric.

\revIIg{For the real SNe~Ia, Figure~\ref{fig:lcpropsvmej} also shows a strong
correlation between \Mej\ and \dmbol{15} ($r = -0.905$):
all $\Mej < \Mch$ SNe~Ia have $\dmbol{15} > 0.75$,
and all SNe~Ia with $\dmbol{15} > 0.95$ have $\Mej < \Mch$.
A matching correlation between \dmbol{15}\ and \dmbol{40}\
($r = 0.929$) captures the same behavior as a geometric invariant of the
bolometric light curve, independent of our interpretation of \dmbol{15}
or \dmbol{40} in terms of \Mej.
The 3-D explosion models shown here all have $\dmbol{15} < 0.75$
(compared with 10/63 real SNe~Ia) and show no clear correspondence
between \dmbol{15}\ and \Mej.  However, \citet{blondin17} find a
\Mej-\dmbol{15} relation in their grid of 1-D NLTE models,
with sub-Chandrasekhar-mass models showing $\dmbol{15} > 0.8$.}

\revIIg{The correlation between \dmbol{15} and \dmbol{40} is intriguing.
Light curve behavior at phase $+15$ days depends upon a rapidly changing,
temperature-dependent optical line scattering opacity,
while at phase $+40$ days it is more stable and driven by gamma-ray opacity.}
\revIIb{The global parameters shaping light curve behavior in both regimes
are \Mej\ and \vKE.  Near maximum light, \citet{arnett82} assumes a constant
gray mean flux opacity $\kappa$ to derive a light curve width timescale
\begin{equation}
\tau_m = \sqrt{ \frac{2 \kappa \Mej}{\beta c \vKE} },
\end{equation}
where $\beta \approx 13.7$ is a dimensionless factor describing the mass
density profile.  At late times, \citet{jeffery99} use a gray mean gamma-ray
scattering opacity to derive the transparency timescale
\begin{equation}
t_0 = \sqrt{ \frac{\kappa_\gamma Q \Mej}{4\pi \vKE^2} },
\end{equation}
where $Q$ is a dimensionless factor depending on the density profile and
\nickel\ distribution.  Taking the ratio we find
\begin{equation}
\tau_m/t_0 = \sqrt{ \frac{\vKE \kappa_\gamma Q}{3 \kappa c} },
\end{equation}
which is independent of \Mej.  In this expression, $c$ is a constant,
$\vKE^{1/2}$ varies by at most 3\% full-scale if $\Mej < \Mch$
(see Figure 10), and $\kappa_\gamma^{1/2}$ is likewise nearly constant in
the limit in which it is applied here \citep{swartz95}.  Among SNe~Ia with
comparable density profiles, \nickel\ distributions, and opacity near
maximum light, the ratio between the diffusion and gamma-ray transparency
timescales should be nearly constant, and $\Mej$ should be the primary
determinant of light curve width.}

\revIIb{\citet{sukhbold18} reach a similar conclusion based on a combination
of semi-analytic arguments (different from those presented above), numerical
experiments, and an empirical analysis of the bolometric light curves from
\citet{scalzo14a}.  However,} \revIIg{Arnett's expression from the
diffusion time hinges on the assumption of constant mean flux opacity.
Interpretation of correlations with quantities like \dmbol{15} or \thp\
may be complicated by the rapidly changing opacity at two weeks after
maximum light, as the ejecta cool and atoms recombine.}

\begin{figure*}
\resizebox{\textwidth}{!}{\includegraphics{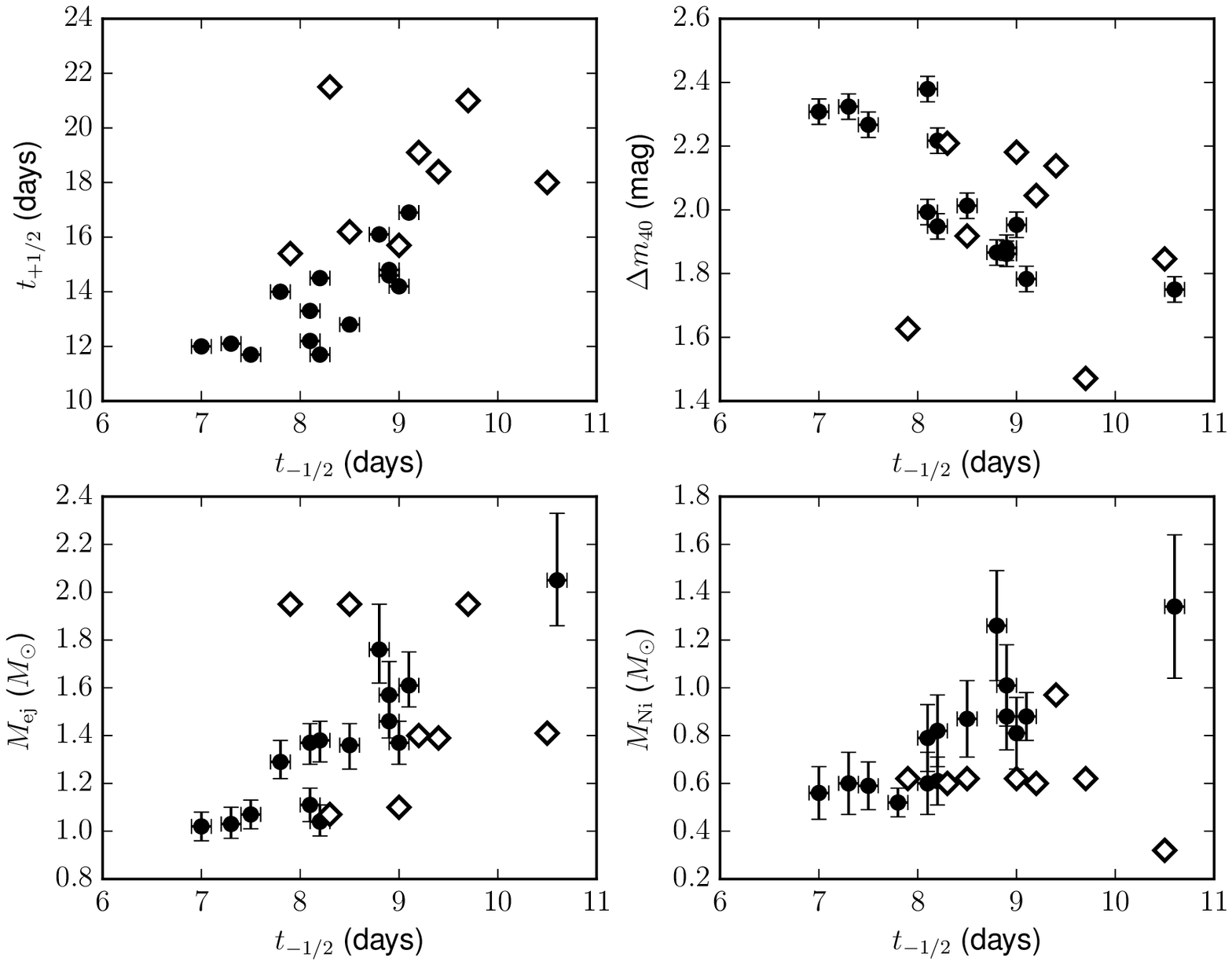}}
\caption{\small Correlations of half-rise time \thm\ with:
half-decline time $\thp$ (top left), \dmbol{40}\ (top right), 
\Mej\ (bottom left), and \MNi\ (bottom right).
Filled circles:  SN~Ia data.  Open diamonds:  explosion models.}
\label{fig:lcpropsvthm}
\end{figure*}

Using the subset of 9~SNe~Ia with good early light curves, we can measure
trends with \thm, which depends on light curve properties in the optically
thick regime, and therefore is a more direct measure of the diffusion time.
Figure~\ref{fig:lcpropsvthm} shows plots of \thm\ against four main quantities
of interest:  \thp, \dmbol{40}, \Mej, and \MNi.  All four quantities show
correlations with \thm.
\revIIg{The explosion models once again either show no correlation or fall
into different regions of parameter space from the real SNe~Ia;
for example, they all have $\thp > 15$~days, while for the real SNe~Ia
$\thp < 15$~days for all but one target.}

Similar relations have been reported before, as early as \citet{contardo00},
but the point has perhaps not received as much attention as it merits.
The correspondences between \thm, \thp, and physical properties of
the explosion suggests that while energy redistribution may affect
the formation of the spectrum or single-band light curves, the total energy
release and shape of the light curve are governed to first order
by the diffusion timescale, rather than by dramatic changes in transparency.

{The details of the very early light curve within the first few days
of explosion may depend upon \revIIg{the spatial distribution of \nickel\
in the outer layers}, where energy can simply escape instead of diffusing
\citep{pn14,pm16}.  Fits of broken power laws to early light curves of some
SNe~Ia show breaks in the power-law index within the first few days after
explosion \citep{zheng13,zheng14,contreras18}.
Where early light curve data are available,
extending our method to include them could provide useful
constraints on the \nickel\ distribution we use to estimate the gamma-ray
transparency of the ejecta at later times.}

\revIIg{Turning briefly to Figure~\ref{fig:lcpropsvmni}, the correlations
between \MNi, \dmbol{40}, and \Lbol{40}\ also reflect the width-luminosity
relation.  We find a weak bolometric width-luminosity relationship
by plotting \Lbol{max}\ vs. \dmbol{15} ($r = -0.526$, $p < 10^{-5}$)
and \MNi\ vs. \dmbol{15} ($r = -0.568$, $p < 10^{-6}$).}


\section{Discussion}
\label{sec:discussion}

Our analysis of a large number of new bolometric light curves are in
agreement with the results of \citet{scalzo14a},
{and replicates the key finding that the ejected mass is closely
linked to both the multi-band and the bolometric light curve shape.}
The added value lies in the diversity of peculiar objects included in our
most recent analysis; in the more rigorous treatment of host galaxy dust
extinction, as inferred from NIR data; and in the \snoopy\ light curve
parametrization, which improves the reliability of bolometric light curve
modeling in the transitional period between the optically-thick and
optically-thin regimes near NIR second maximum.

{We discuss below what we have learned in further detail about
peculiar SN~Ia explosions (\S\ref{subsec:discuss-pecIa}),
the width-luminosity relation (\S\ref{subsec:discuss-wlr}),
and the way forward for more accurate inferences about progenitors
(\S\ref{subsec:discuss-bayes}).}

\subsection{Explosion mechanisms for peculiar SNe~Ia}
\label{subsec:discuss-pecIa}

As previously found by \citet{scalzo12,scalzo14a}, for 1999aa-like and
1991T-like SNe~Ia we infer large ejected masses and large \nickel\ masses.
Many of these luminous, slowly-declining SNe~Ia can be plausibly explained by
Chandrasekhar-mass explosions {that are close to pure detonations.}
A few seem to have moderately super-Chandrasekhar
masses $\Mej > 1.6$~\Msol.  We note that \Mej\ becomes more sensitive
to assumptions about dust extinction, opacities, or distance when
$\MNi/\Mej > 0.7$; below this limit, however, the inferred value of \Mej\
becomes surprisingly robust to these assumptions, depending only on those
factors which most directly influence the light curve shape.  We note that
our modeling assumptions are conservative in the sense that most variations,
as explored in \citet{scalzo14a}, result in comparable or higher \Mej\ for
similar SNe~Ia.  The mass limit for rigidly rotating white dwarfs
is 1.5~\Msol\ \citep{anand65,roxburgh65}, so it may be that most 1991T-like
SNe~Ia share the same explosion mechanism as their spectroscopically normal
counterparts.

Not all SNe~Ia for which we infer $\Mej > \Mch$ are necessarily physically
related to the extreme ``super-Chandra'' SNe~Ia with estimated
$\Mej > 2$~\Msol\ \citep{howell06,scalzo10,silverman11,taub11},
which also have $\MNi > \Mch$ if they are powered exclusively through
radioactive decay {and there are no asymmetries}.
Our work adds SN~2006gz to the list of SNe~Ia with large ejected masses
{inferred} by this technique.  However, several authors
\citep{taub11,hachinger12,taub13a,noebauer16} argue that the extreme
luminosities of super-Chandra SNe~Ia are powered at least partially through
shock interaction of the ejecta from a normal Chandrasekhar-mass explosion
with extended {circumstellar} carbon-oxygen envelopes,
of the kind that might be produced in double-degenerate mergers \citep{it84}.
Our model would not describe the true physics of
such an explosion, but since the required envelopes have masses of order
$0.5~\Msol$, the total ejecta masses are similar to the ones we infer.

Although fast-declining SNe~Ia are likely to have $\Mej < \Mch$,
we find that as a class, peculiar SNe~Ia with cool photospheres are not
necessarily fast-declining.  When the two diverge, the light curve
behavior is a better predictor of \Mej\ than the spectroscopic behavior.
The peculiar SN~2006bt, flagged by \citet{foley10} as an outlier from the
width-luminosity relation for normal SNe~Ia, has a cool photosphere but a
broad light curve, and in our analysis reconstructs with $\Mej > \Mch$,
but with modest \MNi.  SN~2006ot, which \citet{cspdr2} describe
as photometrically similar to SN~2006bt, also shows $\Mej > \Mch$.
SN~2007ba, a fast-declining spectroscopic 1991bg-like SN~Ia,
reconstructs with $\Mej \sim 1.1$~\Msol, still sub-Chandrasekhar but
somewhat higher than other SNe~Ia with similar light curve properties.
There is therefore room for real physical diversity among SNe~Ia with cool
photospheres, and any physical theory for them must explain both
photometric and spectroscopic behavior.

Most interestingly,
\emph{SN~2006bt and SN~2006ot are the first SNe~Ia with normal luminosities
for which super-Chandrasekhar ejected masses have been inferred.}
The membership of SN~2006bt in the Branch-CL subclass is particularly
interesting in light of the 1.66-\Msol\ double-degenerate violent merger model
proposed by \citet{kromer13} to explain the subluminous SN~2010lp, which was
slowly-declining ($\Delta m_{15,B} = 1.24$) for its luminosity.

Together with the extreme ``super-Chandra'' SNe~Ia and the 1991T-like SNe~Ia,
the 2006bt-like SNe~Ia form a third distinct subclass of high-mass explosions.
This suggests diversity in explosion mechanisms for white dwarf systems with
$M > \Mch$ that remains to be fully understood by mapping event classes to
explosion models --- for example, 2006bt-like SNe~Ia to violent mergers,
``super-Chandras'' to explosions inside a carbon-oxygen envelope, and
1991T-like SNe~Ia to single-degenerate pure detonations of spun-up white
dwarfs rotating rigidly near breakup.  The lack of clarity here is perhaps
unsurprising given the theoretical challenges in modeling explosions of
white dwarf systems with $M > \Mch$, especially the effects of magnetic fields
and dynamic merger processes; we hope that our results will spur progress
in this area.

\subsection{Bolometric light curves and the width-luminosity relation}
\label{subsec:discuss-wlr}

The use of the \snoopy\ light curve fitter in this paper is not limited to
improved estimates of host galaxy dust extinction.  The smooth transition
with increasing \sBV\ of the NIR light curves from a single-peaked to a
double-peaked morphology \citep{burns14} improves the reliability of our NIR
corrections between the first and second NIR peaks --- a range of light
curve phases where the transparency of the ejecta is changing rapidly.

The resulting analysis uncovers relationships between \Mej, \revIIg{\MNi,}
and simple morphological properties of the bolometric light curve,
\revIIg{including} a bolometric width-luminosity relation.
Our inference procedure for \Mej\ uses
mostly information contained in \dmbol{40}, and in fact does not actually
use any bolometric light curve points between $B$ maximum and day~$+40$.
There is no \emph{a priori} reason for the ejected mass to correlate strongly
with light curve behavior near day~$+15$, but this is exactly what we find.

While existing numerical explosion models reproduce the trend
between \Mej\ and \dmbol{40}\ reasonably well,
they {are not guaranteed to} reproduce the new
correlations.  No obvious trend between \dmbol{15}\ and \dmbol{40} arises
from the suite of three-dimensional models
shown here \citep{fink10,pakmor12,ruiter13,seitenzahl13a}.
{\citet{blondin17} note that sub-Chandraskehar-mass models are needed
to reproduce such a correlation, in addition to matching any spectroscopic
properties of the resulting SNe~Ia (see their figure~3), although the
bolometric light curves for these models are not yet publicly available
for direct comparison to our results.}

\revIIg{Despite the strong links between \Mej\ and light curve shape,
however, better predictions for \MNi\ or for SN~Ia luminosity remain elusive.}
The bolometric width-luminosity relation
\revIIg{--- \Lbol{max} vs. either \dmbol{15} or \dmbol{40} ---}
is weaker and has more scatter (0.09~dex, or 23\%) than
\revIIg{one might expect} if the link between \Mej\ and \MNi\
were as deterministic as it is in the usual double-detonation scenario
\citep{fink10,sim10,ruiter13}.
Other factors, such as the \emph{distribution} of \nickel\ in the ejecta or
details of radiation transfer in the transition to the optically thin regime,
must also affect the observed luminosity,
as has long been claimed in studies of Chandrasekhar-mass delayed detonations
\citep[e.g.][]{hk96,mazzali07} \revIIg{that go beyond the flux mean opacities
deployed in \citet{arnett82} or \citet{pe00a}.  \sBV\ and similar multi-band
light curve shape parameters must be sensitive to these details.}
Deep understanding of the physical origin
of the width-luminosity relation thus awaits a physical model that can
reproduce both the bolometric and multi-band observed behavior of SNe~Ia.

\subsection{\revIIg{Recent model-theory comparisons}}

We here review some recent developments in comparisons between
theory and observations of SNe~Ia, to put our work in context.

\citet{wygoda17} re-analyze the bolometric light curves of \citet{stritz06}
and \citet{scalzo14a} using their own numerical bolometric light curve code,
inferring \MNi\ and a transparency timescale $t_0$ at which the mean optical
depth of the ejecta to \cobalt\ gamma rays is unity \citep[see][]{jeffery99}.
The parameters as measured from their code are consistent with
\citet{scalzo14a}.  They point out the importance of satisfying both
the bolometric and multi-band width-luminosity relations,
\revIIb{with their results lending support to sub-Chandrasekhar-mass double
detonations and white dwarf collision models}.  The authors
suggest that the most productive way forward is to compare \MNi\ and $t_0$
directly to the output of numerical explosion models rather than to attempt
an inversion, due to the model dependence of many physical details.
\revIIg{While we agree that progress can be made by direct comparison
to simulations,}
\revIIb{we believe an approach like our own also has merit:  in regions of
parameter space populated by multiple models, quantitative estimates of
\emph{a priori} probabilities, and of all known sources of uncertainty,
are required to evaluate which model or scenario is truly the most likely.}

\citet{hoeflich17} consider in great detail the $B$-band and $V$-band
behavior of SNe~Ia, demonstrating that these can be well-represented by
Chandrasekhar-mass delayed detonations with varying central density and
deflagration-to-detonation transition density.  \citet{ashall18} make a
similar claim for fast-declining SNe~Ia, performing abundance tomography for
two individual examples and measuring peak bolometric luminosities for
a larger sample.  Their inferences of \Mej\ for the two examples treated in
detail are point estimates with large ($\sim 25\%$) formal error bars,
consistent with \Mch\ but with little discriminative power.

\citet{blondin17} revisit the Chandrasekhar-mass delayed detonation models
of \citet{blondin13}, while adding new {Chandrasekhar-mass} pulsating
delayed detonation and sub-Chandrasekhar-mass double-detonation models.
They note that the bolometric and multi-band properties of SNe~Ia
with $\Delta m_{15,B} > 1.4$ are much better explained by
sub-Chandrasekhar-mass models than Chandrasekhar-mass models.
\citet{blondin18} compare in detail one of these new sub-Chandrasekhar-mass
explosion models to the 1991bg-like SN~1999by, showing that it fits the
observations better than the nearest competing Chandrasekhar-mass model.
These results stand in sharp contrast to \citet{hoeflich17},
although the \citet{blondin17} sub-Chandrasekhar-mass models still cannot
reach $\Delta m_{15,B} > 1.65$.

\citet{goldstein18} present {a suite of 4500 radiation transfer
simulations based on parametrized ejecta models} and show, in agreement with
\citet{wygoda17} and \citet{blondin17},
that a threshold in decline rate exists below which
only sub-Chandrasekhar-mass models can explain SN~Ia observations.
They point out that the low masses attainable
by the models of \citet{hoeflich17} and \citet{ashall18} may be due in large
part to substantial central cores of stable iron-peak elements in these
models, which are disfavored in 3-D hydrodynamic models but are difficult to
directly rule out observationally.

\subsection{Bayesian inference as a way forward}
\label{subsec:discuss-bayes}

\revIIg{Any interpretive power our technique may have in assigning values
of \Mej\ and \MNi\ to individual SNe~Ia is based on Bayesian inference.}
\revIIb{We believe a Bayesian inversion approach has the potential to go
beyond comparisons of point estimates of parameters for single models.
Without a probability measure on the space of models, including both expert
knowledge \citep[such as those from population synthesis;][]{ruiter13}
and uncertainty sources (including sensitivity to initial conditions),
it is hard to know whether any given model that matches observed parameters
is \emph{a posteriori} either probable or unique.}
\revIIg{Bayesian techniques also, in principle, enable the
fusion of datasets and expert knowledge other than bolometric light curves
(such as spectroscopy or polarimetry) for comparison to
simulations, given appropriate training data.}

The spatial distribution of \nickel\ and of stable iron-peak elements in the
bulk ejecta {are among the main unknowns} making it difficult to reach
consensus on the nature of individual SNe~Ia.  The potential influence of
asymmetries will also need to be treated rigorously to provide the fairest
possible test of explosion models such as violent mergers or white dwarf
collisions.  Neither the radial distribution of stable iron nor the
line of sight towards an asymmetric model can be directly constrained,
{though they can be indirectly constrained through expensive
         observations such as nebular spectra
         \citep{gerardy07,maeda09b,maeda10,maund10}}.
{For the few SNe~Ia that are nearby enough for such observations
         to be obtained, equally expensive forward models are required
         to interpret them.}
{For SNe~Ia too distant for such observations to be acquired,
         stable iron content and line of sight}
must be treated as nuisance parameters in an inversion if inferences for
individual SNe~Ia are to be made.  Only a Bayesian approach allows for
{marginalization over nuisance parameters, or for statistical
combination of heterogeneous types of data in a natural way.}

Given the current state of affairs, semi-analytic calculations have probably
reached the limits of their usefulness.  More information must be brought
to bear on the problem, and the natural progression from our approach would
be to formulate a Bayesian inversion problem on the numerical simulations
themselves.  Although these simulations are expensive, hierarchical emulators
could be built to quickly evaluate an approximation of their output, using
as training data larger suites of simpler radiation transfer models such
as those of \citet{goldstein18}, and then fitting the residuals from more
advanced models like those of \citet{blondin17}.  Such an approach is out of
scope of this work, but may become feasible in the near future.


\section{Conclusions}
\label{sec:conclusions}

We have constructed bolometric light curves for a new sample of well-observed
SNe~Ia located in the Hubble flow, applying the probabilistic inference
framework of \citet{scalzo14a} to estimate their ejected masses and \nickel\
masses.  This more than doubles the number of SNe~Ia in the literature that
have been analyzed by this method.  The basic results of \citet{scalzo14a}
are reproduced using a sample with a different selection function, broader
wavelength coverage, and more rigorous host galaxy dust extinction estimates.
The bolometric light curves are published here for use by the community.

The present work includes a more diverse sample of peculiar SNe~Ia,
including several well-known events and subclass exemplars.
Our analysis yields at least six new SNe~Ia for which $\Mej > \Mch$
at high confidence:
SN~1999dq, SN~2001V, SN~2004gu, SN~2005hj, SN~2006bt, and SN~2006ot.
Our method also {reproduces} the anticipated large ejected mass
of SN~2006gz.  These events are spectroscopically diverse, with a range
of \nickel\ masses from 0.5--1.3~\Msol,
but all have unusually broad light curves.
The large ejected masses of SN~2006bt and SN~2006ot, relative to their
\nickel\ production, explain why and how they deviate from the
width-luminosity relation for normal SNe~Ia, and make them good candidates
for violent merger events.
Contrariwise, cool-photosphere SNe~Ia with fast-declining light curves
consistently present ejected masses less than $\Mch$.
As with normal SNe~Ia, these results suggest that the light curve width is a
better predictor of ejected mass than spectroscopic subtype, although the
details of line formation undoubtedly provide additional information about
the explosion mechanism that is beyond the scope of our analysis.

Finally, correlations between morphological properties of the bolometric
light curve, not reproduced fully by contemporary explosion models, provide
new insight and constraints on explosion models.  The relationships
between {inferred} SN~Ia ejected mass, the radiation diffusion time
in the expanding ejecta, {and multi-band light curve width parameters}
provide further evidence that variation
in ejected mass {is real and has a deep relationship to}
the standardization of SNe~Ia as cosmological candles.  Peculiar SNe~Ia fall
off of this standardization relation to the extent that they violate the
underlying relationship between ejected mass and \nickel\ mass.  Future
development of explosion models focused on understanding the \Mej-\MNi\
relation could both yield insights into the nature of SN~Ia explosions
and provide better standardization relations grounded in basic physics.


\section*{Acknowledgments}

Parts of this research were conducted by the Australian Research Council
Centre of Excellence for All-Sky Astrophysics (CAASTRO), through project
number CE110001020.
RS acknowledges support from ARC Laureate Grant FL0992131.
We thank Ivo Seitenzahl, Brian Schmidt, Robert Kirshner,
Peter Hoeflich, Bruno Leibundgut, and Suhail Dhawan for useful discussions.
The CSP is grateful to the National Science Foundation for continuing support
under grants AST-0306969, AST-0908886, AST-0607438, and AST-1008343.


\bibliography{ms_s15a}{}
\bibliographystyle{mn2e}



\clearpage


\setcounter{section}{0}
\renewcommand{\thesection}{S\arabic{section}}
\setcounter{figure}{0}
\renewcommand{\thefigure}{S\arabic{figure}}


\section{\revIIg{Time-dependent UV Flux Corrections}}
\label{sec:uv-gp}

A suite of \emph{Swift} light curves for 115~SNe~Ia with $z < 0.02$,
15 of which also had published photometry from \csp\ or CfA,
\revIIg{were used for training}.
Of these, 79 had light curves sufficiently well-sampled
(including \emph{Swift} $ubv$ and any optical data) to achieve a convergent
\snoopy\ fit.  For each SN~Ia at each epoch, a UV SED \revIIg{was constructed}
from the \emph{Swift} photometry using the ``best-fit SED'' method of
\citet{brown16}.  This method solves for a self-consistent set of
flux densities in a piecewise-linear SED, constrained to produce synthetic
photometry in the observed filters that best approximate the photometric
observations.  \revIIg{The method naturally accounts for} the red leaks in the
\emph{Swift} \revIIg{UV} filters, and gives results accurate at level of 2\%
of the integrated flux in the range 1600--6000~\AA\ \citep{brown16}.
Each UV-based SED was then de-reddened using the \ebmvhost\ and \rvhost\
parameters from the \snoopy\ fit to produce intrinsic SEDs.

{\citet{milne13} separated \emph{Swift} SNe~Ia into ``NUV-red'' and
``NUV-blue'' subtypes, with \revIIg{normal-velocity \citep{wang09}}
SNe~Ia associated exclusively with the NUV-blue subtype.
\revIIg{\vSi\ might therefore} be a useful
predictor of UV SED behavior for SNe~Ia with no \emph{Swift} observations.}
\revIIg{A search of the Open Supernova Catalog \citep{guillochon16} yielded
25~SNe~Ia with both \emph{Swift} coverage and optical spectra within 3~days
of $B$-band maximum light.}
From these spectra, \vSi\ was measured using the technique
described in \citet{scalzo12}, in which a pseudo-continuum is estimated
by applying a Savitzsky-Golay filter to the data, and uncertainties on the
position of the line minimum are \revIIg{estimated} by Monte Carlo
\revIIg{simulation}.

\begin{figure}
\resizebox{0.47\textwidth}{!}{\includegraphics{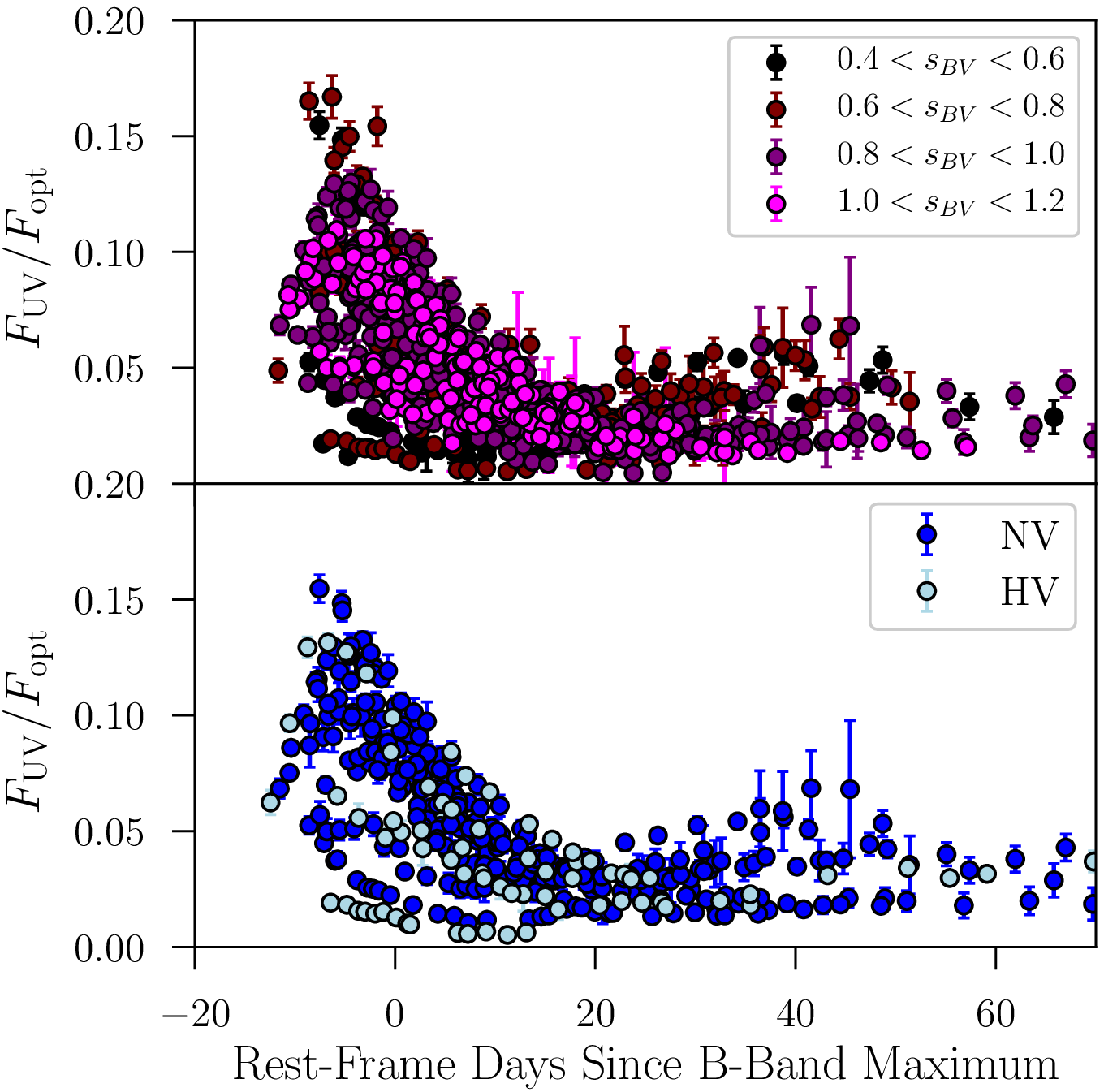}}
\caption{\small Ratio of UV (1600--3300~\AA) to optical (3300--8500~\AA)
flux as a function of $B$-band light curve phase.  Top:  79~SNe~Ia in four
bins of light curve shape --- $0.4 < \sBV < 0.6$ (black),
$0.6 < \sBV < 0.8$ (maroon), $0.8 < \sBV < 1.0$ (purple), and
$1.0 < \sBV < 1.2$ (magenta).  Bottom:  25~SNe~Ia divided between the
\protect\citet{wang09} normal (dark blue) and high-velocity (light blue)
subtypes.  {No clear correlation is seen of UV flux fraction with \sBV\ or
Wang type is seen.}}
\label{fig:uv-frac}
\end{figure}

Figure~\ref{fig:uv-frac} shows the ratio of UV to optical flux as a function
of rest-frame light curve phase for the training set.  No clear correlation
of UV flux fraction with \sBV\ or with {the Wang spectral} type is
seen; examples can be found for any light curve shape or velocity spanning
nearly the full range of UV behavior.

\revIIg{To build a \emph{time-dependent} template for}
UV flux, we normalize each SED
to unit flux density at 4355~\AA\ (the $B$-band point) and take the logarithm.
We then train a GP model to predict the (log) flux density based on phase,
wavelength, and \sBV, much like the NIR template correction used by
\citet{scalzo14a}; the covariance is
\begin{equation}
k_\mathrm{3D}(\vec{x}, \vec{x}') =
   \exp\left[(\vec{x} - \vec{x}')^T \vec{\Theta} (\vec{x} - \vec{x}') \right],
\end{equation}
with feature vector $x = (t, \sBV, \log(\lambda))$ and hyperparameters
$\vec{\Theta} = \mathrm{diag}(\Theta_t, \Theta_\sBV, \Theta_\lambda)$.
We choose to predict the UV SED, instead of a correction as a fraction of
the optical flux, in order to capture the covariance between inferred
UV flux fraction and extinction properties.  This procedure ensures that
uncertainties in reddening are treated self-consistently in
downstream modeling.

\begin{figure*}
\resizebox{\textwidth}{!}{\includegraphics{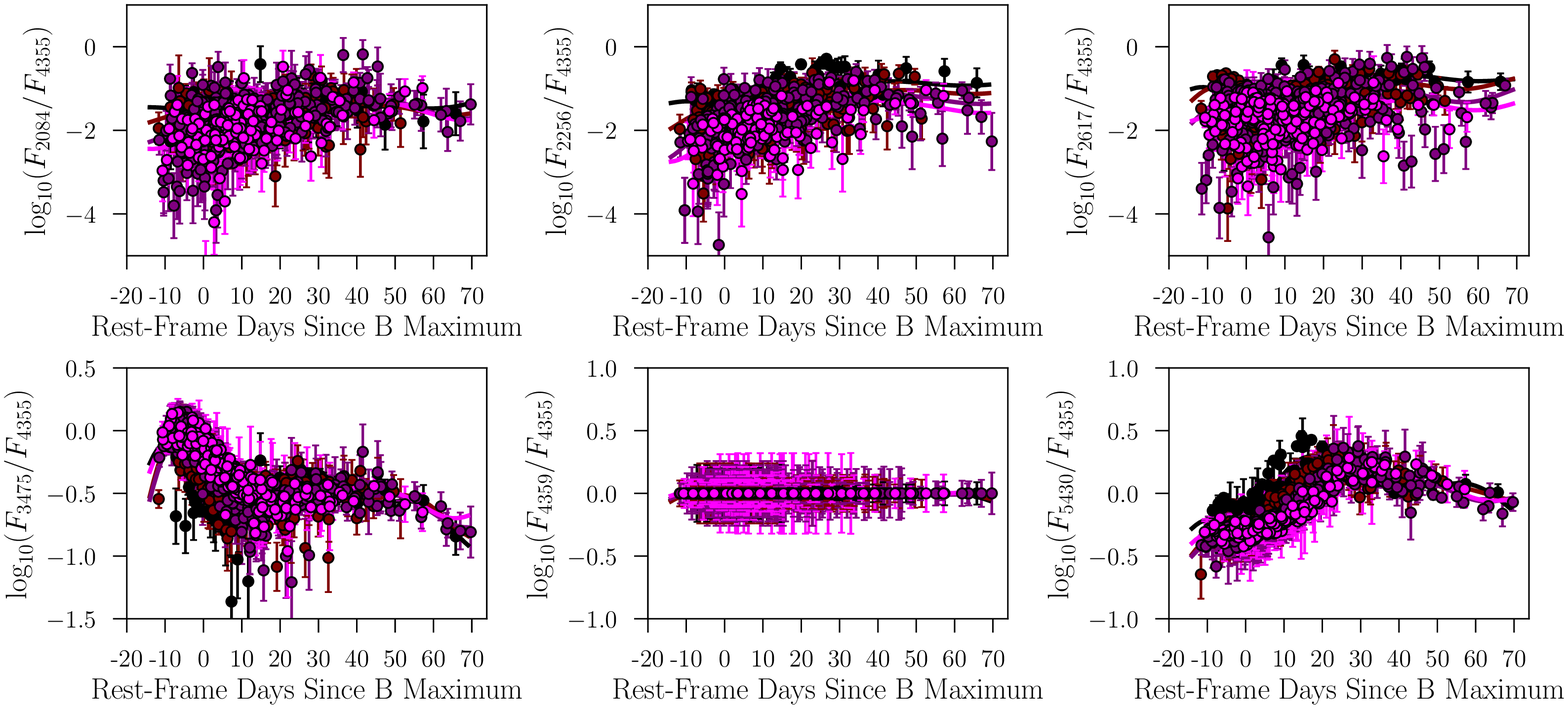}}
\caption{\small Log flux ratios at different wavelengths for SEDs built from
\emph{Swift} data.  Colors show different ranges of \sBV\ as in the top panel
of Figure~\ref{fig:uv-frac}.  Bold lines with cross-hatches show Gaussian
process fits, with 68\% confidence regions, evaluated at the midpoint of each
bin in the \sBV\ sequence.}
\label{fig:uv-gp}
\end{figure*}

The resulting fitted template is shown in Figure~\ref{fig:uv-gp}.
While \sBV\ is not a particularly good predictor of near-UV color,
the predicted $B$-band flux density at given \sBV\ provides a convenient
normalization for the SED.  The {\emph{Swift}}
$u-b$ and $b-v$ colors show the usual tight
dispersion with light curve shape.  For the bluer bands, the dispersion is
much larger ($\sim 0.4$~dex RMS); this full dispersion is propagated through
to the final uncertainty estimation on the integrated bolometric flux.


\section{\revIIg{Priors for Explosion Property Inference}}
\label{sec:priors}

\subsection{\revIIg{Fiducial priors}}
\label{subsec:priors-runF}

Our fiducial modeling assumptions correspond to ``Run~F''
of \citet{scalzo14a}, which successfully reproduce the ejected masses of
a diverse suite of three-dimensional numerical explosion models:
sub-Chandrasekhar-mass double detonations \citep{fink10,ruiter13},
super-Chandrasekhar-mass violent mergers \citep{pakmor12}, and
Chandrasekhar-mass delayed detonations \citep{seitenzahl13a}.  Run~F assumes:
\begin{enumerate}
\item an ejecta density profile \mbox{$\rho(v) \propto [1 + (v/\vKE)^3]^{-3}$};
\item no dense core of stable iron-peak elements;
\item less than 0.05~\Msol\ of unburned material; and
\item $\alpha = 1.0$.
\end{enumerate}
These assumptions are difficult or impossible to fully confirm observationally
with current techniques, but all are motivated by one or more contemporary
explosion models from different groups, and the effects of varying them are
described in detail in \citet{scalzo14a}.  The density profile is the most
influential factor, with different functional forms changing the absolute
inferred mass scale up or down by about $0.1$~\Msol.
The form factor $\alpha$ affects \MNi\ directly, and affects \Mej\ through
the inferred transparency of the ejecta to \cobalt\ gamma rays.
The fraction of unburned material and the presence or absence of a stable
iron-peak core turn out to be less influential, affecting \Mej\ and \MNi\
at the 0.05~\Msol\ level or less.
\revIIb{The Run~F results are close to the median
of the distribution of \Mej\ for each SN~Ia obtained by varying over the eight
different priors explored in \citet{scalzo14a}.}

\revIIb{Thus, sensitivity to priors introduces some uncertainty on the overall
mass scale, as well as in the posterior variance of \Mej\ for individual
SNe~Ia, which may affect the interpretation of our present work.  However,
the range of \Mej\ values inferred for normal SNe~Ia was not found to be
sensitive to the choice of prior in \citet{scalzo14a}.  The existence of a
correlation between \Mej\ and \salt\ $x_1$ is also robust:  a change of prior
alters the fitting formula's intercept but has little effect on its slope.}

\revIIg{\citet{scalzo14c} apply such a fitting formula to a larger sample of
337 SNe~Ia from the Supernova Legacy Survey \citep{betoule14} to infer the
intrinsic distribution of \Mej, using it to make separate arguments that
Run~F is well-calibrated.  The \Mej\ distribution shows a sharp peak at
1.4~Msol, which is well-explained by Chandrasekhar-mass models, but with a
tail extending down to 0.8~\Msol.  Any choice of prior producing results
very different from Run~F would imply a peak at some other mass scale,
which would require explanation by some other scenario.  A peak near
1.2~\Msol\ could be explained by violent white dwarf mergers
\citep{ruiter13}, though this would result in very few Chandrasekhar-mass
SNe~Ia to satisfy nucleosynthetic constraints \citep{seitenzahl13b}.
Similarly, \citet{ptk14} note that the white dwarf collisions of
\citet{kushnir13} would imply a peak in \Mej\ around 1.6~\Msol, given the
luminosity distribution of these events as compared with normal SNe~Ia.}

The overall shape of the distribution \revIIg{
(in particular, the width of the peak or the presence of a low-mass tail)}
will be affected only by strong systematic variations in density or
maximum-light opacity as a function of decline rate.
\revIIg{To address this possibility, the present work explores}
two other effects not previously
considered in \citet{scalzo14a}:  more sophisticated priors on $\alpha$
which take into account correlations with other global parameters not
accounted for in a simple Gaussian prior, and the possible influence of
unobserved mid-infrared (MIR) flux on the bolometric light curve.


\subsection{Model-dependent covariances between $\alpha$
            and other global explosion parameters}
\label{subsec:priors-alpha}

The true distribution of $\alpha$ is uncertain and reflects dependence
on tuning to particular suites of numerical models;
the details of radiation transport near maximum light are too complex to
unfold fully in a semi-analytic model and must be simulated numerically.
For the semianalytic models of \citet{arnett82},
$\alpha$ is exactly equal to 1, by construction.
For the 1-D radiation hydrodynamic models of \citet{hk96}, $\alpha$
varied in the range 0.6--1.4 depending on the explosion mechanism.
{
The 1-D models of \citet{blondin13,blondin17} assume homologous expansion
after initial burning ceases, but treat radiation transfer in non-local
thermodynamic equilibrium (NLTE), leading to $\alpha$ in the range 0.9--1.3.}
To capture possible covariances between $\alpha$ and global explosion
parameters \revIIg{that are lost by using a fixed value of $\alpha$ or a
simple Gaussian prior, we construct empirical priors from the published
model grids of \citet{hk96} and \citet{blondin13,blondin17} as GP regressions,
and simulate them for comparison with the fiducial Run~F prior.}

\begin{figure}
\resizebox{0.47\textwidth}{!}{\includegraphics{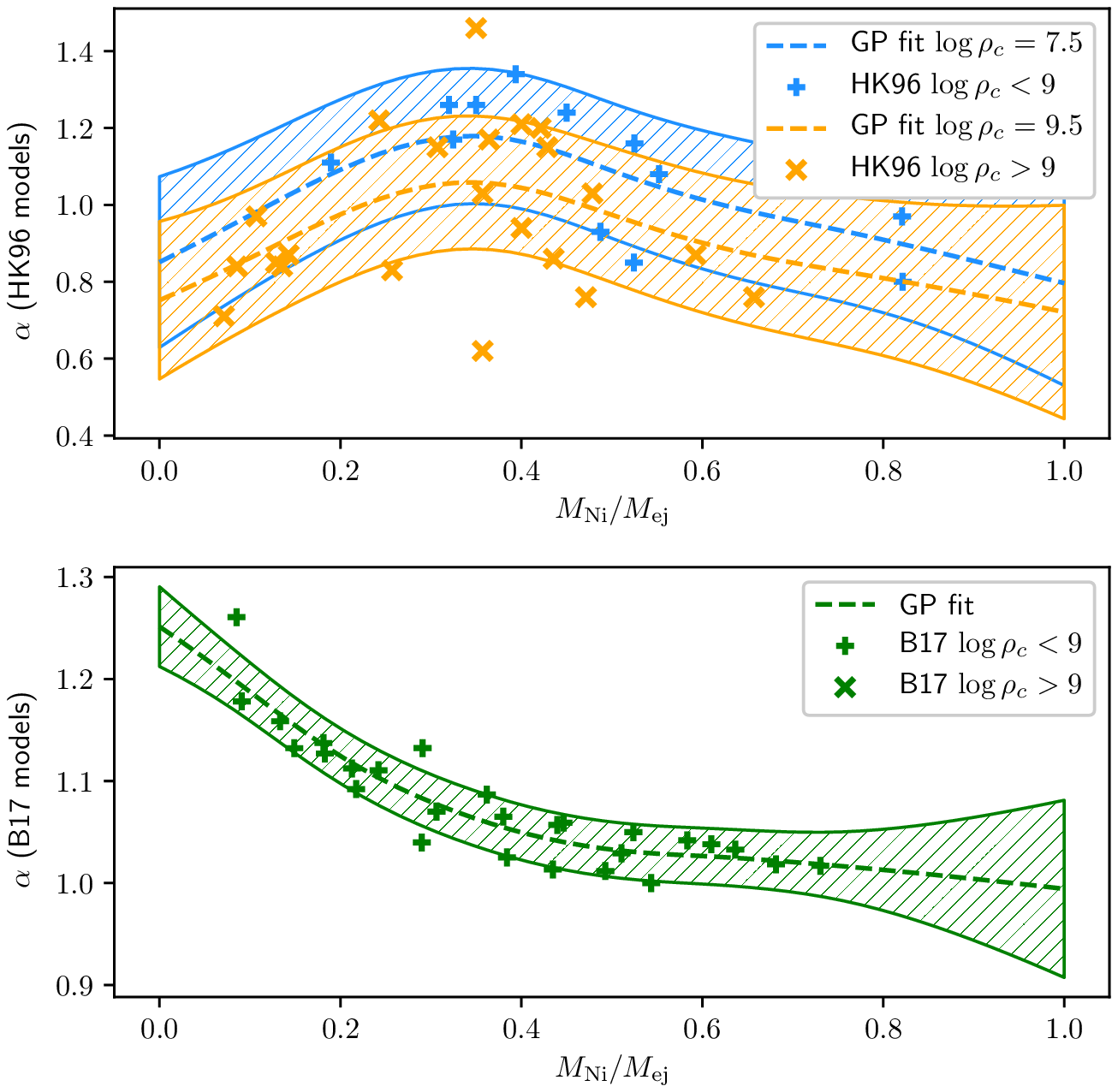}}
\caption{\small Variation in the form factor $\alpha$ for the 1-D models
of \protect\citet{hk96} (top) and \protect\citet{blondin17} (bottom).
Low-density explosion
models are marked with vertical crosses, and high-density explosion models by
diagonal crosses.  Mean GP predictions are shown as dotted lines, while the
hatched regions are 68\% confidence intervals.
{These regressions are used as priors linking $\alpha$ to
\MNi\ and \Mej\ in our modeling.}
}
\label{fig:alpha-gp-prior}
\end{figure}

\revIIg{The new priors use} the \nickel\ mass ratio $\MNi/\Mej$ and the
white dwarf central density $\rho_c$ as potential predictors of $\alpha$,
\revIIb{on the basis that both parameters affect the opacity (via temperature)
and the radial distribution of \nickel.}  Since the intrinsic dispersion
of $\alpha$ around the mean trend for each explosion model grid
is unknown, we fit for it as a hyperparameter.  The results for both
{explosion model grids} are shown in Figure~\ref{fig:alpha-gp-prior}.

The GP regression for the \citet{hk96} model grid predicts $\alpha$ with
RMS dispersion of 17\%.  The input data comprise two clusters in central
density:  a high-density cluster of Chandrasekhar-mass delayed detonations
with \mbox{$\log_{10} \rho_c \sim 9.3$--9.5},
and a low-density cluster of sub-Chandrasekhar-mass double detonations and
super-Chandrasekhar-mass tamped detonations
with \mbox{$\log_{10} \rho_c \sim 7.0$--7.6}.
The low-density cluster has a higher mean $\alpha$ (1.10) than the
high-density cluster (0.97).  There is also a trend with \nickel\ mass,
with a maximum in $\alpha$ being attained around $\MNi/\Mej \sim 0.35$.

In contrast, the GP regression for the \citet{blondin13,blondin17} model grid
has a much smaller dispersion (2.5\%), and $\alpha$ is determined entirely by
$\MNi/\Mej$.  The trend with $\MNi/\Mej$ is also monotonically decreasing.
As in the \citet{hk96} model grid, delayed detonations, pulsating delayed
detonations, and sub-Chandrasekhar-mass detonations all contribute.
\citet{blondin13} note that the input hydrodynamic models are produced by
the same code, so that the difference probably lies in different treatments
of opacity and its effects on the optical depth and diffusion time.

These two contrasting cases demonstrate one of the challenges in assigning
a more complex prior on $\alpha$, namely, that different codes run on
similar physical scenarios may obtain different results; it can be hard
to tell whether regularities among models in a grid reflect true physical
constraints or merely result from approximations made in each code.
As the physical relationship between $\alpha$ and global physical parameters
is clarified further by future work, we can expect the uncertainty in our
inferences to decrease.  For the time being, {because each informative
prior encodes specific physics assumptions that may help to constrain the
final results, we prefer to treat each one separately in order to be
transparent about those assumptions, rather than using a
single broad prior insensitive to underlying correlations between model
parameters.}


\subsection{Potential influence of unobserved MIR flux}
\label{subsec:priors-mir}

Unobserved MIR flux is difficult to correct for, since few MIR
observations of SNe~Ia exist from which a template could be constructed,
and since theoretical predictions of MIR emission \revIIg{remain uncertain}.
Rayleigh-Jeans extrapolations of
our own SEDs suggest that unobserved MIR flux as a fraction of observed flux
from 1600--17500~\AA\ ranges from less than 5\% near maximum light to around
10\% near day~+40, with no strong dependence on decline rate \sBV.
\revIIb{As discussed above,} the flux difference between maximum light and
later phases is the most effective predictor of \Mej\ within our modeling
framework, so uncaptured flux at late times could {bias}
our ejected mass estimates.  We cross-check this finding against empirical
SEDs from data in the literature and a theoretical ansatz.

\citet{johansson17} provide MIR light curves of 9 normal SNe~Ia with decline
rates $0.9 < \Delta m_{15,B} < 1.3$ measured with the \emph{Spitzer}
space telescope; except for the well-observed SN~2014J, there are in general
very few light curve points available for light-curve phases between maximum
light and day~$+80$.  We used available light curve data to build SEDs at
phases relevant to our analysis for two SNe~Ia observed
by \citet{johansson17}:  SN~2007le ($\Delta m_{15,B} = 1.10$~mag)
and SN~2009ig ($\Delta m_{15,B} = 0.89$~mag).
Integrated flux in the range 2--6~microns made up 5\% of total bolometric
flux for SN~2007le at day $+66$, and only 3\% for SN~2009ig at day $+36$,
in each case less than the Rayleigh-Jeans extrapolation predicts.
These two cases suggest that the effects of neglecting MIR flux are minimal
{compared to other systematic sources of uncertainty from factors such
as distance, reddening, and model inadequacy,}
at least for SNe~Ia on the slow-declining side of normal.

In Chandrasekhar-mass delayed detonation models of faster-declining SNe~Ia,
MIR contributions could be larger than this (Hoeflich, private communication).
As an additional test, two new analyses were run in which
the late-time bolometric flux was boosted by a linearly interpolated fraction
between day~$+55$ and day~$+85$, based on models for normal SNe~Ia
(``case~A'':  10\% increase at day~$+60$, 25\% increase at day~$+80$)
and for 1991bg-like SNe~Ia
(``case~B'':  25\% increase at day~$+60$, 40\% increase at day~$+80$).
We caution that this ansatz applies only to one explosion scenario, and that
it does not capture correlations between unobserved MIR flux and other
explosion parameters.
Future publications of synthetic bolometric light curves from numerical
explosion models, {as well as new MIR observations of SNe~Ia,}
will enable us to examine these effects in more detail.


\section{\revIIg{Comments on Individual Supernovae}}
\label{sec:sncomments}


\subsection{SN~1999aa}
\label{subsec:sncomments-1999aa}

This CfA SN~Ia from \citet{jha07} is the exemplar of its own slowly-declining
subclass \citep{li01,strolger02,scp99aa,silverman12a}, intermediate between
Branch-normal SNe~Ia and 1991T-like SNe~Ia.  A fit to the multi-band light
curve using \salt\ yields $x_1 = +1.17$, $c = -0.08 \pm 0.02$,
while \snoopy\ gives $\sBV = 1.15$ and negligible host galaxy reddening
($\ebmvhost = 0.005 \pm 0.016$~{mag}).  Its CMB-centric redshift is
$z_\mathrm{CMB} = 0.01522$, giving $\mu = 34.10 \pm 0.15$~{mag}
for our assumed
cosmology and peculiar velocity systematic.  Given these constraints,
SN~1999aa reconstructs as a moderately bright Chandrasekhar-mass event
($\Mej = 1.42^{+0.10}_{-0.07}$~\Msol, $\MNi = 0.78^{+0.16}_{-0.13}$~\Msol).


\subsection{SN~1999dq}
\label{subsec:sncomments-1999dq}

This CfA SN~Ia from \citet{jha07} is slow-declining
($x_1 = +0.83$, $\sBV = 1.20$), with a 1999aa-like
spectrum \citep{silverman12b} and significant reddening
($\ebmvhost = 0.154$~{mag}, $\rvhost = 3.5$~{mag}),
putting its corrected intrinsic color at
$(B-V)_\mathrm{max} = -0.12$~{mag}.  Its CMB-centric
redshift is $z_\mathrm{CMB} = 0.0136$, leading to a distance modulus
$\mu = 33.89 \pm 0.15$~{mag} for our assumed cosmology and peculiar
velocity systematic.  SN~1999dq reconstructs as a luminous
super-Chandrasekhar-mass event at $> 95\%$ confidence
($\Mej = 1.67^{+0.19}_{-0.14}$~\Msol, $\MNi = 1.28^{+0.25}_{-0.22}$~\Msol).

\citet{neill09} measure $\MNi = 0.96 \pm 0.09$~\Msol\ for SN~1999dq
(assuming zero host galaxy extinction) via the technique of \citet{howell09},
which uses the \citet{hsiao07} spectroscopic template for the bolometric
correction near maximum light.  Correcting this zero-extinction estimate
using the observed color as a proxy for reddening by dust in the host galaxy,
\citet{neill09} measure $\MNi = 1.22 \pm 0.11$~\Msol\ for SN~1999dq,
the highest in their sample, and comparable to our own estimate.


\subsection{SN~2001V}
\label{subsec:sncomments-2001V}

This CfA SN~Ia from \citet{jha07} was noted by \citet{vinko03} as being
slow-declining and exceptionally luminous, with estimated $M_B \sim -20$
and a 1999aa-like spectrum near maximum light.  The light curve width
parameters from our fits are $x_1 = 0.85$ and
$\sBV = 1.19$.  The CfA light curve is exceptionally complete and suggests
a very blue SN~Ia.  Our inferred $A_V = 0.46 \pm 0.14$~{mag} is
non-negligible, while \citet{mandel11} infer much stricter limits
$A_V < 0.11$~{mag} (68\% confidence).  Our extinction estimate implies
$(B-V)_\mathrm{max} = -0.19$~{mag}, $M_B = -20.21$~{mag}.
However, even with no host galaxy extinction correction,
$M_B = -19.72$~{mag}
for SN~2001V, indicating a large mass of \nickel\ ($> 0.9~\Msol$) without
accounting for flux bluewards of $B$-band which can contribute substantially
to the luminosity \citep{scalzo14b}.  Using the \snoopy\ extinction values,
SN~2001V reconstructs as super-Chandrasekhar-mass
($\Mej = 1.78^{+0.20}_{-0.15}$~\Msol, $\MNi = 1.29^{+0.25}_{-0.22}$~\Msol).
Using the \bayessn\ upper limit,
SN~2001V is still super-Chandrasekhar-mass at $> 95$\% confidence
($\Mej = 1.60^{+0.14}_{-0.10}$~\Msol, $\MNi = 0.97^{+0.18}_{-0.16}$~\Msol).


\subsection{SN~2004gu}
\label{subsec:sncomments-2004gu}

This slowly-declining SN from \citep{cspdr1}, with a broad light curve
($x_1 = 1.37$, $\sBV = 1.19$) and photometric and
spectroscopic similarities to SN~2006gz, has been mentioned in
several papers about extreme ``super-Chandra'' SNe~Ia
\citep{yuan10,taub11,silverman11}.  A fit with \snoopy\ suggests substantial
host galaxy extinction ($\ebmvhost = 0.18 \pm 0.02$~{mag}) with a
fairly shallow extinction law ($\rvhost = 1.8 \pm 0.3$).  The \bolomass\
reconstruction shows it to be super-Chandrasekhar-mass at $>95\%$~confidence
($\Mej = 1.55^{+0.12}_{-0.08}$~\Msol, $\MNi = 0.91^{+0.13}_{-0.12}$~\Msol),
similar to SN~2005hj and the 1999aa-like SNF~20070506-006 \citep{scalzo14a}.


\subsection{SN~2005hj}
\label{subsec:sncomments-2005hj}

\citet{qhw07} discussed this SN~Ia in the context of its slow
\ion{Si}{2}~$\lambda6355$ velocity evolution near maximum light.
They attribute this behavior to the presence of a dense shell in the outer
layers of ejecta, a property of several non-homologous explosion models
such as the tamped detonation and pulsating delayed detonation scenarios
\citep{kmh93,hk96}.  Similar slow evolution has also been observed in
the super-Chandrasekhar-mass SN~2007if \citep{scalzo10} and in several
super-Chandrasekhar-mass candidates with 1991T-like spectra \citep{scalzo12},
and in fact occur frequently in 1991T-like SNe~Ia \citep{benetti05}.
\citet{scalzo12} argued that these explosions were tamped detonations
resulting from prompt double-degenerate mergers exploding
inside compact carbon-oxygen envelopes, and used \vSi\ to infer the
contributions of these envelopes to the total ejected mass.

SN~2005hj itself is a slowly-declining
($x_1 = +1.43$, $\sBV = 1.19$), 1999aa-like \citep{qhw07} SN~Ia with modest
reddening $\ebmvhost = 0.12 \pm 0.02$~{mag}, $\rvhost = 1.4 \pm 0.5$.
It reconstructs as super-Chandrasekhar-mass at $>95$\%~confidence
($\Mej = 1.55^{+0.12}_{-0.08}$~\Msol, $\MNi = 0.89^{+0.12}_{-0.11}$~\Msol),
similar to SN~2004gu and the 1999aa-like SNF~20070506-006 \citep{scalzo14a}.
Although the presence of a shell in the ejecta could affect the intrinsic
color of the SN and hence the inferred reddening by host galaxy dust,
\MNi\ is not large enough to significantly affect our estimate of \Mej,
as for SN~2005ls (see below).


\subsection{SN~2005ls}
\label{subsec:sncomments-2005ls}

This CfA SN~Ia from \citet{cfa3} is as slowly-declining
($x_1 = +0.83$, $\sBV = 1.24$) as spectroscopic 1991T-like SNe~Ia,
but its earliest spectrum was taken at day~$+8$ (CBET 324), so its
spectroscopic behavior before that point is unclear.  It has a red color
(\salt\ $c = +0.29$) and a large inferred
$\ebmvhost = 0.41 \pm 0.02$~{mag}
($\rvhost = 2.8 \pm 0.2$), so its inferred intrinsic color
$(B-V)_\mathrm{max} = -0.12$ is blue but not extremely blue.  Our inferred
$A_V = 1.15 \pm 0.10$~{mag} is also consistent with the value
obtained by \citet{mandel11} ($A_V = 0.81^{+0.93}_{-0.68}$~{mag}),
though the latter infers lower $\rvhost = 2.1 \pm 0.2$.

Using the \snoopy\ extinction values, SN~2005ls reconstructs as
a luminous super-Chandrasekhar-mass candidate
($\Mej = 1.72^{+0.16}_{-0.12}$~\Msol, $\MNi = 1.30^{+0.19}_{-0.17}$~\Msol).
Using the looser \code{bayessn} constraints on reddening,
SN~2005ls remains luminous but is consistent with being Chandrasekhar-mass
($\Mej = 1.47^{+0.12}_{-0.09}$~\Msol, $\MNi = 1.01^{+0.18}_{-0.17}$~\Msol).

This suggests that uncertainty in the extinction, and particularly in the
extinction law slope, can in some cases drive systematic variations in our
inference of \Mej; the sampled models are ones with the minimum allowed
fractions of intermediate-mass elements, so that adding \nickel\ requires
a larger ejected mass.  Although SN~2005ls formally passes our selection
criteria, it has only one set of observations near $B$-band maximum light,
with the next available observations at day~$+10$; this may make the host
galaxy extinction parameters more susceptible to systematic errors
(e.g., in $K$-corrections within \snoopy) than other SNe.

While SN~2005ls may thus be a good candidate for a super-Chandrasekhar-mass
explosion, its status is dependent upon input assumptions to a greater extent
than for other SNe~Ia in this work.  In other respects, SN~2005ls is generally
consistent with the behavior of other 1991T-like SNe~Ia in our sample that
may be modestly, but not extremely, super-Chandrasekhar-mass.


\subsection{SN~2006bt}
\label{subsec:sncomments-2006bt}

This SN~Ia is described as peculiar by \citet{foley10} and
\citet{cspdr2}:  it is a Branch-CL event
(with a cool photosphere near maximum light and \ion{Ti}{2}~$\lambda4100$
features similar to SN~1991bg), but has a relatively broad light curve
(\salt\ $x_1 = 0.07 \pm 0.10$)
and a red apparent color (\salt\ $c = 0.16 \pm 0.01$).
\citet{foley10} reports that the MLCS2k2 light curve fitter \citep{jha07}
estimates $A_V = 0.43 \pm 0.05$~{mag}, even though the SN is far
from the centre
of an early-type galaxy, has non-standard color curves and shows no sign of
\ion{Na}{1}~D absorption; they believe the actual extinction is negligible.
Due to its photometric peculiarities, SN~2006bt is excluded from the
training set for the \snoopy\ color model of \citet{burns14}.  If applied
anyway to the CfA light curve, \snoopy\ gives $\sBV = 1.19$,
$\ebmvhost = 0.26 \pm 0.02$~{mag},
$\rvhost = 2.6 \pm 0.8$~{mag}, suggesting a
highly extinguished event.
Fitting the \csp\ light curve gives similar results.

We analyze the CfA light curve of SN~2006bt assuming no host galaxy reddening,
although we still use $\sBV = 1.19$ to characterize the light curve and to
evaluate the mean function for the GP interpolation.  We find SN~2006bt is
super-Chandrasekhar-mass at $> 95$\% confidence, but with a moderate mass of
\nickel\ ($\Mej = 1.62^{+0.23}_{-0.11}$~\Msol,
          $\MNi = 0.48^{+0.08}_{-0.06}$~\Msol).
This may be an underestimate of \Mej\ if SN~2006bt's late-time NIR behavior
is similar to SN~2006ot's (see below), which seems likely given the similarity
between their light curves where data are available \citep{cspdr2}.

The \csp\ light curve for SN~2006bt does not formally pass our selection
criteria, with the last optical-wavelength observations taken at day~$+39$,
and the last NIR observation at day $+11$; the coverage between that point
and maximum is excellent.  If we relax the requirements slightly, using
the day~$+39$ point as indicative of the late-time light curve behavior,
and assume no host galaxy reddening, \bolomass\ predicts
\mbox{$\Mej = 1.82^{+0.37}_{-0.17}$~\Msol} and
\mbox{$\MNi = 0.48^{+0.08}_{-0.06}$~\Msol},
consistent with the estimate from the CfA light curve.

In contrast to SN~2005ls or other highly reddened events, SN~2006bt's modest
value of \MNi\ makes its inferred value of \Mej\ more robust to systematic
errors in reddening or distance.  A cross-check using a loose Gaussian
prior $\ebmvhost = 0.1 \pm 0.1$~{mag},
$\rvhost = 3.1 \pm 1.0$ suggests that
$\Mej > 1.5~\Msol$ (99\% confidence) for a range of plausible values of
\ebmvhost\ and \rvhost.

It thus seems likely either that SN~2006bt is super-Chandrasekhar-mass,
with $\Mej \sim 1.7~\Msol$, or that it somehow violates one or more of our
modeling assumptions (such as stratified ejecta or a standard density profile)
without showing dramatic spectroscopic peculiarities.
In either case, typical Chandrasekhar-mass models are probably not
well-suited to describe the physics of the explosion.


\subsection{SN~2006gt}
\label{subsec:sncomments-2006gt}

This SN~Ia is one of two Branch-CL events for which \snid\ finds a
1991bg-like spectroscopic subtype.  It is a fast-declining event
($x_1 = -1.81$, $\sBV = 0.53$) with a red color
$(B-V)_\mathrm{max} = 0.05$~{mag} after correction for mean extinction.
\snoopy\ estimates of the host galaxy extinction using the \csp\ and CfA
light curves are mutually consistent:
$\ebmvhost = 0.06 \pm 0.02$~{mag} (CfA: $0.08 \pm 0.04${mag}),
$\rvhost = 2.7 \pm 1.0$ (CfA: $3.3 \pm 1.3$).
We use the \csp\ light curve to reconstruct it as a sub-Chandrasekhar-mass
event ($\Mej = 0.86^{+0.05}_{-0.03}$~\Msol,
       $\MNi = 0.27^{+0.04}_{-0.04}$~\Msol).


\subsection{SN~2006gz}
\label{subsec:sncomments-2006gz}

This peculiar SN~Ia from \citet{hicken07} is one of the exemplars of the
extremely luminous ``super-Chandra'' events with unusually slowly-declining
light curves \citep[see also][]{howell06,scalzo10,yuan10,silverman11,taub11}.
However, discussion of the super-Chandrasekhar-mass nature of this event has
hinged mostly on inferences about \MNi, which has been called into question
\citep{maeda09,maeda09b,scalzo10}.
\citet{hicken07} also cite evidence for \ion{C}{2}
absorption at early phases, although such features are now believed to be
present in 20\%--30\% of normal SNe~Ia \citep{thomas11,parrent11,folatelli12}.
\citet{taub13a} argue that the nebular spectra of SN~2006gz are more similar
to the super-Chandra SNe~Ia 2007if and 2009dc than to 1991T-like SNe~Ia,
which presumably can be explained as near-Chandrasekhar-mass explosions.

A fit with \salt\ yields $x_1 = 2.18 \pm 0.12$, $c = 0.11 \pm 0.03$, while
a \snoopy\ fit suggests $\sBV = 1.31$, the largest value in our sample.
The host galaxy extinction parameters according to \snoopy\ are
$\ebmvhost = 0.17 \pm 0.02$~{mag} and
$\rvhost = 4.1 \pm 1.9$, consistent with
the values assumed by \citet{hicken07}.  Under these assumptions, we infer
$\MNi = 1.34^{+0.31}_{-0.28}$~\Msol, again consistent with \citet{hicken07},
and $\Mej = 2.05^{+0.28}_{-0.19}$, comparable to the value inferred
for SN~2007if using similar methods \citep{scalzo10}.

Given the danger that overcorrection for host galaxy reddening may bias
\MNi, and hence \Mej, we run a separate reconstruction assuming zero
host galaxy extinction.  This run yields $\Mej = 1.93^{+0.41}_{-0.17}$~\Msol,
$\MNi = 0.64^{+0.11}_{-0.09}$~\Msol, so that the super-Chandrasekhar-mass nature
of SN~2006gz is secure irrespective of the \nickel\ content.  We can place
a firm 99\% confidence lower limit $\Mej > 1.6~\Msol$ on the ejecta even
with the most conservative assumptions about reddening.


\subsection{SN~2006ot}
\label{subsec:sncomments-2006ot}

\citet{cspdr2} note the photometric peculiarity of SN~2006ot as well as its
similarity to SN~2006bt, although spectroscopically it is a Branch-BL SN~Ia
with the highest at-maximum \ion{Si}{2} velocity in our sample
\citep[13989 \kms;][]{folatelli13}.  As with SN~2006bt above, SN~2006ot was
excluded from the training set for the \snoopy\ color model of
\citet{burns14}, and \citet{cspdr2} find both objects suffered little to no
host galaxy extinction.

SN~2006ot does not pass our full selection criteria, with the first
optical-wavelength data appearing at day~$+3$ after $B$-band maximum.
However, it has well-sampled coverage in \csp\ $BVgriYJ$ past day~$+40$ and
\csp\ $u$ out to day $+35$, so despite some uncertainty on the maximum-light
bolometric luminosity, our reliance on templates to fill in regions of
missing observations is minimal.  This is fortunate, since we find that
\dmbol{40}\ is overestimated by 0.1~{mag}
if we use the \snoopy\ predictions
for the $YJH$ light curves instead of the GP interpolation model.
Given that SN~2006bt does not have late-time $u$ or NIR data, we can be
reasonably certain that we have underestimated its mass somewhat by using
template corrections for this missing flux near day~$+40$.

As in the case of SN~2006bt, we analyze SN~2006ot assuming negligible host
galaxy extinction ($\ebmvhost = 0.00 \pm 0.05$~{mag}).
We propagate an uncertainty of $\pm 3$~days on the date of $B$-band maximum
into the rise time, and adjust the maximum-light bolometric luminosity
upwards by 10\% to reflect the decline by day $+3$.  The resulting \bolomass\
run shows that SN~2006ot is super-Chandrasekhar-mass at high confidence
($\Mej = 1.91^{+0.43}_{-0.20}$, $\MNi = 0.50^{+0.09}_{-0.07}$).
A separate analysis with loose Gaussian priors
$\ebmvhost = 0.1 \pm 0.1$~{mag}
and $\rvhost = 3.1 \pm 1.0$ places a firm 99\%
confidence level lower limit of 1.6~\Msol\ on \Mej\ given any plausible
value for \ebmvhost.  This lower limit is conservative, in the sense that
all adjustments made above tend to reduce the inferred value of \Mej.


\subsection{SN~2007ba}
\label{subsec:sncomments-2007ba}

This SN~Ia is the other spectroscopic 1991bg-like in our sample.
Like SN~2006gt, it is also a fast-declining event
($x_1 = -1.64$, $\sBV = 0.59$) with a red color
$(B-V)_\mathrm{max} = 0.05$~{mag} after correction for mean extinction.
\snoopy\ estimates of the host galaxy extinction using the \csp\ and CfA
light curves are marginally consistent:
$\ebmvhost = 0.23 \pm 0.02$~{mag} (CfA: $0.16 \pm 0.04$~{mag}),
$\rvhost = 1.7 \pm 0.3$ (CfA: $1.8 \pm 0.9$).  Using the \csp\ light curve,
we infer a sub-Chandrasekhar mass for SN~2007ba at $> 95\%$ confidence
(\mbox{$\Mej = 1.11^{+0.09}_{-0.07}$~\Msol},
 \mbox{$\MNi = 0.48^{+0.08}_{-0.07}$~\Msol}).

The high mass, relative to other Branch-CL events with similar light curve
shapes, is intriguing, and worth investigating for systematic errors.
SN~2007ba has its last $u$-band measurement at day~$+22$, although this
should be late enough for the $u$-band color to have stabilized and for later
measurements to be reasonably well-predicted by \snoopy.
SN~2007ba also has excellent NIR coverage, with $YJH$ measurements extending
out to day~$+48$.

With a maximum \emph{a posteriori} probability of $P_\mathrm{fit} = 0.004$,
the \bolomass\ model fit is not particularly good, further suggesting
SN~2007ba is unique among the objects considered in our study.
Inspection of the light curve fit shows that the radioactive deposition curve
agrees well with the observed bolometric light curve at days~$+44$ and $+46$,
but underpredicts day~$+55$ by almost 25\%.  The $B$-band light curve falls
below magnitude 21.0 near day~$+40$ and begins to \emph{brighten} after
day~$+45$, suggesting that measurements at such faint magnitudes may not be
reliable despite larger reported uncertainties.  If we remove the day~$+55$
measurement from the fit and run \bolomass\ again,
we get nearly the same results
(\mbox{$\Mej = 1.09^{+0.07}_{-0.06}$~\Msol},
 \mbox{$\MNi = 0.48^{+0.08}_{-0.07}$~\Msol}) with a much better fit
($P_\mathrm{fit} = 0.338$).  We also get the same results if we allow points
as early as day~$+30$ into the fit, arguing that less massive ejecta should
enter the Compton-thin regime (assumed by \bolomass) earlier than most SNe~Ia.

\end{document}